\documentclass[twoside,12pt]{elsarticle}
\usepackage{epsfig}
\usepackage{amsmath,amssymb,appendix}
\usepackage[bookmarks=true,colorlinks,breaklinks,%
            citecolor=pink,linkcolor=pink,anchorcolor=pink,filecolor=pink,urlcolor=pink]{hyperref}
\allowdisplaybreaks
\journal{Progress in Particle and Nuclear Physics}

\def\P{\vec P}

\newcommand{\be}{\begin{equation}}
\newcommand{\ee}{\end{equation}}
\newcommand{\bea}{\begin{eqnarray}}
\newcommand{\eea}{\end{eqnarray}}

\begin{document}

\title{ \vspace{1cm} Low-Energy Heavy-Ion Reactions and the Skyrme Effective Interaction}
\author{P.\ D.\ Stevenson$^1$ and M. C. Barton$^1$
\\
$^1$Department of Physics, University of Surrey, Guildford, Surrey, GU2 7XH, United Kingdom}
\begin{abstract}
  The Skyrme effective interaction, with its multitude of parameterisations, along with its implementation using the static and time-dependent density functional (TDHF) formalism have allowed for a range of microscopic calculations of low-energy heavy-ion collisions.  These calculations allow variation of the effective interaction along with an interpretation of the results of this variation informed by a comparison to experimental data.  Initial progress in implementing TDHF for heavy-ion collisions necessarily used many approximations in the geometry or the interaction.  Over the last decade or so, the implementations have overcome all restrictions, and studies have begun to be made where details of the effective interaction are being probed.  This review surveys these studies in low energy heavy-ion reactions, finding significant effects on observables from the form of the spin-orbit interaction, the use of the tensor force, and the inclusion of time-odd terms in the density functional.
\end{abstract}
\maketitle
\tableofcontents
\section{Introduction}
Heavy-ion collisions combine the rich dynamics of a many-body out-of-equilibrium open quantum system with the complexities of the residual part of the strong interaction which leaks out of the small, but neither fundamental or point-like, nucleons, causing them to stick loosely together some of the time, and to fall apart at others.  Understanding heavy-ion reactions across all energy scales is necessary to understand stellar nucleosynthesis \cite{Bennett2012}, the synthesis of superheavy nuclei \cite{Oganessian2015,Umar2014}, the properties of nuclear matter \cite{Li2008,Suraud1989,Lynch2009}, the QCD phase diagram \cite{Foka2016,Braun-Munzinger2016} as well as the understanding of reaction mechanisms themselves \cite{Hinde2015,Khuyagbaatar2015,Hinde2002,mandaglio,Simenel2012a}.  

Among the theoretical techniques used to study heavy-ion reactions, methods based on time-dependent Hartree-Fock have recently achieved the status of having sufficiently mature implementations free of limiting approximations, and running at a suitable speed, such that systematically varying the effective interaction in the calculations is possible.  It is such studies that form the main subject of the present review.  The practical implementations, using the Skyrme interaction, are in some sense parameter-free, in that one has a framework using an effective interaction fitted to ground state data and nuclear matter properties, with no further adjustment to dynamics.  Structure and reaction effects are together determined self-consistently from the interaction, subject to the approximations of the mean-field and one gives no further adjustment.  In another sense, the variation among the sets of available effective interactions are parameters of the calculations.  We attempt to summarise here what has been learnt from exploring different Skyrme force parameterisations within low-energy heavy-ion reaction calculations.

Overlapping this subject area are other recent review articles, to which the reader is referred:  A review in which extensive coverage of theoretical approaches to dynamics of heavy-ion collisions in TDHF and its extensions is presented by Simenel and Umar \cite{Simenel2018}.  This review extensively covers the detail of the calculational framework, which we cover in less detail here, instead concentrating more on the role of the effective interaction.  Spin-dependent aspects of the effective interaction and their role in heavy-ion reactions at low and higher energy have recently been reviewed by Xu et al. \cite{Xu2015}.  Recent developments in experimental studies of heavy-ion fusion reactions are covered by Back et al. \cite{Back2014}.

The border between the kind of calculations we have included in this review, and those not, is a somewhat arbitrary choice.  Using other theoretical approaches such as transport theory \cite{Bertsch1988,Baran2005,Buss2012}, suitable for higher energy collisions (above a few hundred MeV/A), also required the use of an effective interaction, varying which produces different outcomes that can be compared with nature.  We concentrate on the mean-field + Skyrme approach as a lowest order, and self-consistent, first step to address the role of the effective interaction in low-energy heavy-ion collisions.

The review is laid out as follows:  We give a brief summary of the TDHF approach, noting the availability of recent detailed reviews, in section \ref{sec:theory}.  Section \ref{sec:skyrme} covers in some detail the Skyrme effective interaction, and its implementation in time-dependent mean-field approaches.  The range of available works in which aspects of the effective interaction are systematically studied is surveyed in section \ref{sec:tdhfresults}.

\section{Theoretical methods} \label{sec:theory}
\subsection{Time-dependent Hartree-Fock}
The Time-dependent Hartree-Fock (TDHF) method, as originally posited by Dirac \cite{Dirac1930}, is the basic microscopic quantal approximation to nuclear dynamics with effective nucleon-nucleon interactions \cite{Koonin1980,Davies1985,Negele1982,Simenel2008,Simenel2012,Nakatsukasa2016,Simenel2018}.  It can be derived as a truncation of the hierarchy of dynamical equations which couple together all many-body density matrices, limiting to the one-body density matrix, and assuming that the two-body density can be expressed as an antisymmetrised product of one-body matrices \cite{Koonin1980}.  Alternatively, the TDHF equations can be derived from the principle of least action within a space of Slater Determinant wave functions \cite{Simenel2012}, or from a more general variational principle in which both the state of the system and the desired observable are optimised, with TDHF arising as the result when the expectation value of one-body observables are optimised.  This more general variational principle is due to Balian and V\'en\'eroni \cite{Balian1981}.

One derivation for the TDHF equations, following references \cite{Villars1972,Koonin1975}, begins from the time-dependent Schr\"odinger equation
\begin{equation}\label{eq:tdse}
  i\hbar\frac{\partial}{\partial t}|\Psi(t)\rangle = \hat{H}|\Psi(t)\rangle.
\end{equation}
One then considers the time-evolution of the one-body density matrix
\be
\rho_{\beta\alpha} = \langle\Psi(t)|a^\dagger_\alpha a_\beta|\Psi(t)\rangle
\ee
as
\be
i\hbar\frac{\partial\rho_{\beta\alpha}}{\partial t} = \left(i\hbar\frac{\partial}{\partial t}\langle\Psi|\right)a^\dagger_\alpha a_\beta|\Psi\rangle + \langle \Psi|a^\dagger_\alpha a_\beta\left(i\hbar\frac{\partial}{\partial t}|\Psi\rangle\right).
\ee
From (\ref{eq:tdse}), its adjoint, and the Hermiticity of $\hat{H}$, the time-derivative of the one-body density matrix becomes
\be \label{eq:drhodt}
i\hbar\dot{\rho}_{\beta\alpha} = \langle\Psi|\left[a^\dagger_\alpha a_\beta,\hat{H}\right]|\Psi\rangle,
\ee
using the dot to notate a time-derivative.

Now, one supposes a Hamiltonian of the form
\be \label{eq:supposehamil}
\hat{H}=\hat{T}+\hat{V}=\sum_{\alpha\beta}t_{\alpha\beta}a^\dagger_\alpha a_\beta + \frac{1}{2}\sum_{\alpha\beta\gamma\delta}V_{\alpha\beta\gamma\delta}a^\dagger_\alpha a^\dagger_\beta a_\delta a_\gamma
\ee
where the kinetic energy is
\be
t_{\alpha\beta} = \langle\alpha|-\frac{\hbar^2\nabla^2}{2m}|\beta\rangle
\ee
and
\be
V_{\alpha\beta\gamma\delta}=\langle\alpha\beta|V|\gamma\delta\rangle
\ee
are the two-body interaction matrix elements.

Using (\ref{eq:supposehamil}) in (\ref{eq:drhodt}) and the anticommutation relationships for fermion creation and annihilation operators gives
\be \label{eq:obde}
\begin{aligned}
i\hbar\dot{\rho}_{\beta\alpha} &= \sum_\delta\left(t_{\beta\delta}\rho_{\delta\alpha}-\rho_{\beta\delta}t_{\delta\alpha}\right)\\
&+\frac{1}{2}\sum_{\delta\lambda\sigma}\left\{\left(V_{\beta\delta\lambda\sigma}-V_{\beta\delta\sigma\lambda}\right)\rho^{(2)}_{\lambda\sigma\delta\alpha}+\rho^{(2)}_{\beta\lambda\delta\sigma}\left(V_{\sigma\delta\lambda\alpha}-V_{\delta\sigma\lambda\alpha}\right)\right\}
\end{aligned}
\ee
where the two-body density matrix is
\be
\rho^{(2)}_{\lambda\sigma\delta\alpha}=\langle\Psi|a^\dagger_\alpha a^\dagger_\delta a_\sigma a_\lambda|\Psi\rangle.
\ee

The equation of the time-evolution of the one-body density matrix, (\ref{eq:obde}), thus links the one-body density matrix to the two-body density matrix via the two-body interaction.  Similarly, if one follows the same procedure, higher-order equations couple together each successive $N$-body density matrix, leading to the BBGKY hierarchy.

To truncate the hierarchy and retrieve the TDHF equations, the two-body density matrix is approximated as
\be
\rho^{(2)}_{\lambda\sigma\delta\alpha} = \rho_{\sigma\delta}\rho_{\lambda\alpha}-\rho_{\sigma\alpha}\rho_{\lambda\delta}.
\ee
Substituting this into (\ref{eq:obde}) and defining the one-body (Hartree-Fock) potential as
\be
W_{\alpha\beta} = \sum_{\delta\sigma}(V_{\alpha\delta\beta\sigma}-V_{\alpha\delta\sigma\beta})\rho_{\sigma\delta}
\ee
gives
\begin{equation}
  \label{eq:obde2}
  \begin{aligned}
    i\hbar\dot{\rho}_{\beta\alpha} &= \sum_\delta\left((t_{\beta\delta}+W_{\beta\delta})\rho_{\delta\alpha}-\rho_{\beta\delta}(t_{\delta\alpha}+W_{\delta\alpha})\right)\\
    &= \sum_\delta\left(h_{\beta\delta}\rho_{\delta\alpha}-\rho_{\delta\beta}h_{\delta\alpha}\right)
    \end{aligned}
 \end{equation}
with
\be
h_{\alpha\beta}=t_{\alpha\beta}+W_{\alpha\beta}.
\ee

In shorthand, one can then write the compact form of the TDHF equations as
\begin{equation}\label{eq:tdhf}
  i\hbar\dot\rho=[h,\rho].
\end{equation}

Practical implementations of TDHF work in a representation of single particle states that make up the Slater Determinant wave function
\be
|\Psi\rangle = \left(\prod_{i=1}^Aa^\dagger_i\right)|0\rangle
\ee
where $a^\dagger_i$ creates a particle in state $i$ and is given by
\begin{equation}
  a^\dagger_i = \int d\alpha a^\dagger_\alpha\langle\alpha|\psi_i\rangle.
\end{equation}
One can then show \cite{Koonin1975} that the TDHF equation (\ref{eq:tdhf}) can be satisfied if each $|\psi_i\rangle$ evolves in time according to
\begin{equation}\label{eq:dpsidt}
  i\frac{d}{dt}|\psi_i\rangle = h|\psi_i\rangle.
\end{equation}
In practice, one works in a coordinate representation;
\be
\psi_i(\pmb{r}s\tau) = \langle \pmb{r}s\tau|\psi_i\rangle
\ee
and solves the equation (\ref{eq:dpsidt}) by time-evolution of initial wave functions in small increments of time.  Details of practical numerical solution of the TDHF equations can be found elsewhere \cite{Simenel2008,Kim1997}, including implementations in which the full code is published \cite{Maruhn2014,Schuetrumpf2018}.  We mention also that the closely-allied time-dependent relativistic mean field has been implemented with published code \cite{Berghammer1995}, which is restricted to collective motion of a single nucleus, such as the case of giant resonances, but not set up for the calculation of heavy-ion collisions.  Results using this code have been presented in which the external field is used to directly simulate Coulomb excitation as if from a projectile \cite{Vretenar1993} but in lieu of calculations with varied interactions that can be compared to the wider literature we do not include it subsequently in the discussion.  Earlier implementations of time-dependent relativistic mean-field have been reported   \cite{Bai1987} in which a brief indication of TDHF-like behaviour is made before concentrating on relativistic energies beyond the scope of this review, and as an exemplar for the density-constrained TDHF method \cite{Cusson1985}.  The existing time-dependent relativistic mean field codes are implemented in the so-called {\textit no-sea approximation} in which states in the Dirac sea are ignored, and it is suggested \cite{Ring2012} that a full (and technically-challenging) implementation of the Dirac sea is needed for the study of dynamics within the relativistic energy density functional / mean-field approach.

\subsection{Heavy-ion reactions in TDHF}
In order to describe a heavy-ion reaction in TDHF, one must start with a suitably-prepared initial condition.  This is usually two nuclei in their ground states, calculated with a particular effective interaction.  The two nuclei are placed in a computational box in coordinate space, such that the wave functions from each nucleus do not overlap (or barely overlap and are re-orthogonalised) and combined into a single Slater Determinant.  Each single particle wave function is then given a Galilean boost such that nucleus $1$ is moving with momentum $\pmb{P}_1$ and nucleus 2 with momentum $\pmb{P}_2$.  The initialisation process can be written \cite{Maruhn2014}
\begin{equation}
\begin{array}{ll}
   \psi_{\alpha,1}(\pmb{r},s;t\!=\!0)
    =e^{i\pmb{p}_1\cdot\pmb{r}}
    \psi_{\alpha,1}^{\mbox{(stat)}}(\pmb{r}-\pmb{R}_1,s),&
      \pmb{p}_1 =\frac{\pmb{P}_1}{A_1}
    \quad,
    \\
    \psi_{\alpha,2}(\pmb{r},s;t\!=\!0)
    = e^{i\pmb{p}_2\cdot\pmb{r}}
    \psi_{\alpha,2}^{\mbox{(stat)}}(\pmb{r}-\pmb{R}_2,s),&
    \pmb{p}_2=\frac{\pmb{P}_2}{A_2}
    \quad.
 \label{eq:inittdhf}
\end{array}
\end{equation}
which gives the transformation from the stationary solutions, indicated by $\psi^\mathrm{(stat)}$, shifted to $\pmb{R}_1$ and $\pmb{R}_2$ and boosted by $\pmb{p}_1$ and $\pmb{p}_2$.  $A_1$ and $A_2$ are the mass numbers of the two nuclei.  $\pmb{P}_1$ and $\pmb{P}_2$ are set up so that $\pmb{P}_1=-\pmb{P}_2$.  One typically specifies a total centre of mass energy for a collision, along with an impact parameter and appropriate values for the initial momenta are calculated assuming a Rutherford trajectory from infinity to the initial nuclear placement on the grid.

\begin{figure}[tb]
\begin{center}
\begin{minipage}[t]{10 cm}
\includegraphics[width=10cm]{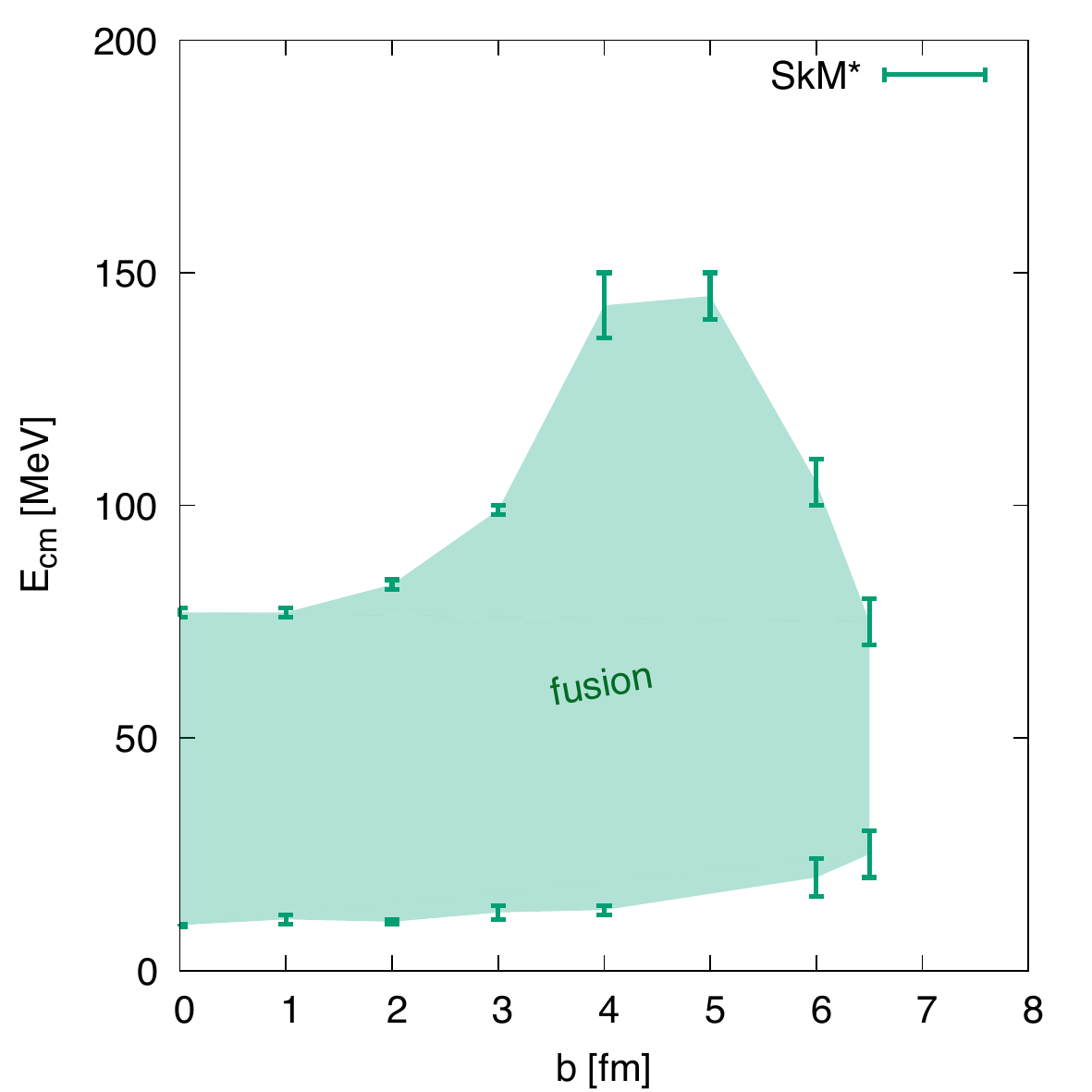}
\end{minipage}
\begin{minipage}[t]{16.5 cm}
  \caption{Map of the fusion region (shaded) as a function of centre-of-mass energy $E_{CM}$ and impact parameter $b$ for collisions of $^{16}$O + $^{40}$O using the SkM* Skyrme interaction \cite{Bartel1982}.  Error bars show the energy intervals in which the transition between fusion and not-fusion is found.
    \label{fig:window_skms}}
\end{minipage}
\end{center}
\end{figure}

Following a collision, the final state of the system can be analysed.  The results of a single TDHF calculation give a final state in which different channels are mixed.  Further interpretation can require post-processing, e.g. in the form of projection onto good quantum numbers \cite{Simenel2010,Sekizawa2013}.  An accessible outcome of a standard TDHF calculation is whether the collision resulted in fusion or not-fusion.  In the first case, one must run the calculation long enough to see that following the collision, the compound nucleus undergoes at least one full oscillation of the internal motion without separation into fragments.  It can still be the case that later in the calculation fission might occur, but it gives an adequate operational definition of fusion.

Reactions in which fusion does not take place result in more than one fragment in the final state.  In this case, the reaction may be a below-barrier approach with Coulomb excitation, a grazing reaction, transfer, fusion-fission, quasi-fission, a deep-inelastic collision, or a mixture of a combination of these.

Figure \ref{fig:window_skms} shows the region of the $E_\mathrm{CM}$--$b$ plane in which fusion occurs for $^{16}$O+$^{16}$O calculations using the SkM* \cite{Bartel1982} interaction, giving one a typical idea of the fusion landscape that arises in TDHF calculations in terms of the regions of fusion and not-fusion.

From such calculations, one can extract a fusion cross-section based on a sharp-cutoff formula \cite{Blair1954,Bonche1978,Davies1985} arising from the fact that in TDHF at a given energy and impact parameter the probability of fusion is either $0$ or $1$:

\begin{equation} \label{eq:xsec}
  \begin{aligned}
    \sigma_f &= \frac{\pi}{k^2}\sum_l(2l+1)\\
    &= \frac{\pi\hbar^2}{2\mu E_{\mathrm{CM}}}\left[(l_>+1)^2-(l_<+1)^2\right]\\
    &\approx \pi(b_>^2-b_<^2).
    \end{aligned}
  \end{equation}
Here, $\mu$ is the reduced mass of the dinuclear system, $E_\mathrm{CM}$ is the centre of mass energy, $l_<$ is the minimum angular momentum at which fusion occurs and $l_>$ the maximum angular momentum at which fusion occurs at the given energy. $b_<$ and $b_>$ are corresponding minimum and maximum impact parameters. The approximate equality in (\ref{eq:xsec}) comes from taking the quantised angular momentum over to a semi-classical limit as a function of the continuous variable $b$.  Examples of such calculations for specific effective interactions are shown in section \ref{sec:tdhfresults}.  From the calculations leading to Figure \ref{fig:window_skms} one sometimes reduces the information by characterising the upper and lower lines of the locus delineating fusion and not-fusion for comparison between different interactions \cite{Umar1986,Stevenson2015a,Stevenson2015}. 

\subsection{Frozen HF approximation}
Without invoking the full complexity of a TDHF calculation, one can bring information form the effective interaction to bear using methods designed to extract a nucleus-nucleus (NN) potential from the microscopic interaction \cite{Brink1975,Washiyama2008,Umar2009}.  In particular, one can begin from static Hartree-Fock ground state calculations and make use of the so-called frozen Hartree-Fock approximation.  One uses the ground-state densities from Hartree-Fock calculations to generate a nucleus-nucleus (NN) potential and defines the nuclear part of the NN potential as \cite{Denisov2002,Brueckners}
\be
V(\pmb{R}) = E(\pmb{R}) - E_{\mathrm{HF}}[\rho_1]-E_{\mathrm{HF}}[\rho_2]
\ee
in which $\pmb{R}$ is the radius vector between the two nuclei, and $E_{\mathrm{HF}}[\rho_1]$ and $E_{\mathrm{HF}}[\rho_1]$ are the Hartree-Fock energies for the nuclei with given densities $\rho_1$ and $\rho_2$.  These are defined for the Skyrme interaction in the next section (\ref{eq:functional}), but may be written schematically as
\begin{equation}
  E_{\mathrm{HF}}[\rho_j] = \int \mathcal{E}[\rho_j(\pmb{r})]d\pmb{r}.
\end{equation}
in which $\mathcal{E}(\pmb{r})$ is an energy density functional.

The total interaction energy is defined in terms of the same functional as
\begin{equation}
  E(\pmb{r}) = \int \mathcal{E}[\rho_1(\pmb{r})+\rho_2(\pmb{R}-\pmb{r})]d\pmb{r}.
\end{equation}

From the NN potential, one can read off the barrier height (the maximum in the potential) or and use as input for two-body scattering or fusion calculations, with e.g. a coupled-channels method \cite{Hagino2012}.  

\subsection{Density-Constrained TDHF}\label{sec:dctdhf}
An improvement of the Frozen Hartree-Fock approximation involves allowing the densities of the incoming nuclei to change as a function of separation distance to account for the Pauli exclusion principle as the nuclei begin to overlap \cite{Simenel2017}.  The principal approach along these lines is the Density-Constrained Time-Dependent Hartree-Fock approach \cite{Cusson1985,Umar2006a,Oberacker2015,Simenel2013,Umar2014}.

In DC-TDHF, the densities are computed by a single TDHF calculation at an energy above the Coulomb barrier.  At each point along the trajectory, a density-constrained Hartree-Fock calculation is performed to find the energy of a nucleus with the given density but without the internal excitations associated with the TDHF calculation.  One then extracts a NN potential in which effects such as necking, shape changes, re-ordering of single-particle states, and the Pauli principle are taken into account.  From the potential one can solve a two-body Schr\"odinger equation with incoming wave boundary conditions \cite{Rawitscher1964} to obtain interaction cross sections.  The complexity of a coupled-channel calculation is not needed as the DC-TDHF potential implicitly includes exited state information.

\section{Skyrme and Skyrme-like Interactions} \label{sec:skyrme}
\subsection{The Skyrme Interaction}
The Skyrme interaction was suggested by its eponymous proposer as an effective two- and three-body interaction for use in the independent particle model \cite{Skyrme1959}\footnote{A four-body term was also proposed in the original paper, though this has not become a standard feature of implementations of the Skyrme interaction}.  A link between it and more realistic interactions can be made by, for example, the density-matrix expansion method of Negele and Vautherin \cite{Negele1972}, which implicitly makes the link via nuclear matter, or alternatively with more direct approaches \cite{Dobaczewski2016,RuizArriola2016}.  One can re-formulate the Skyrme interaction as a energy density functional (EDF) \cite{Bender2003,Jones2015}.   The EDF formalism is strictly the correct way to approach the problem for irreducibly density-dependent versions of the Skyrme interaction \cite{Erler2010}.  However, here we use the language of the interaction as the starting point for the derivation of the EDF since it is the basis of most available comparisons of the underlying forces in heavy-ion collisions within this mean-field framework.

The original Skyrme interaction may be written as a potential as \cite{Skyrme1959,Bender2003,Erler2011,Barton2018}
\begin{equation} \label{eq:twobodyforce}
V = \sum_{i<j} \upsilon_{ij}^{(2)} + \sum_{i<j<k} \upsilon_{ijk}^{(3)}.
\end{equation}
The two and three body Skyrme interactions, in a form essentially the same as that originally given, can be written as
\begin{equation} \label{eq:skyv12}
\begin{aligned}
\upsilon_{12}^{(2)}  &= \ t_{0} (1+x_{0}P_{\sigma}) \delta ( \pmb{r}_{1} - \pmb{r}_{2} ) \ + \ \dfrac{t_{1}}{2} \Big( 1 + x_{1} P_{\pmb{\sigma}} \Big) \Big[\delta ( \pmb{r}_{1} - \pmb{r}_{2} ) \pmb{k}^{2} + \pmb{k}'^{2} \delta ( \pmb{r}_{1} - \pmb{r}_{2} ) \Big] \\& + \ t_{2} \Big( 1 + x_{2} P_{\pmb{\sigma}} \Big) \pmb{k}' \cdot \delta ( \pmb{r}_{1} - \pmb{r}_{2} ) \pmb{k} \ + \ i W_{0}(\pmb{\sigma}_{1} + \pmb{\sigma}_{2}) \cdot \Big( \pmb{k}' \times \delta ( \pmb{r}_{1} - \pmb{r}_{2} ) \pmb{k} \Big) \\&
+ \dfrac{t_{e}}{2} \bigg\{ \big[ 3 (\pmb{\sigma}_{1} \cdot \pmb{k}')(\pmb{\sigma}_{2} \cdot \pmb{k}') - (\pmb{\sigma}_{1} \cdot \pmb{\sigma}_{2}) \pmb{k}'^{2} \big]\delta ( \pmb{r}_{1} - \pmb{r}_{2}  ) \\&
 + \big[ 3 (\pmb{\sigma}_{1} \cdot \pmb{k})(\pmb{\sigma}_{2} \cdot \pmb{k}) - (\pmb{\sigma}_{1} \cdot \pmb{\sigma}_{2}) \pmb{k}^{2} \big]\delta ( \pmb{r}_{1} - \pmb{r}_{2} ) \bigg\} \\& 
+ t_{o} \bigg\{ \big[ 3 (\pmb{\sigma}_{1} \cdot \pmb{k}')(\pmb{\sigma}_{2} \cdot \pmb{k}) - (\pmb{\sigma}_{1} \cdot \pmb{\sigma}_{2}) \pmb{k}' \cdot \pmb{k} \big] \delta ( \pmb{r}_{1} - \pmb{r}_{2}  ) \bigg\} \\
\end{aligned}
\end{equation}
and
\begin{equation}
\upsilon_{123}^{(3)} \ = \ t_{3} \delta(\pmb{r}_{1}-\pmb{r}_{2}) \delta(\pmb{r}_{2} - \pmb{r}_{3} ) 
\end{equation}
respectively.
Here $\pmb{\sigma}$ are Pauli spin matrices, $\pmb{k} \ = \ \dfrac{1}{2i} (\pmb{\nabla}_{1} - \pmb{\nabla}_{2})$ acting to the right,  and $\pmb{k}' \ = \ - \dfrac{1}{2i} (\pmb{\nabla}_{1}' - \pmb{\nabla}_{2}')$, acting to the left.

Included in (\ref{eq:twobodyforce}) are undetermined constants which are associated with the contact term ($t_0$, $x_0$), momentum-dependent terms ($t_1$, $t_2$, $x_1$, $x_2$), the spin-orbit term ($W_0$) \cite{Bell1956}, tensor terms ($t_e$ and $t_o$), and a three-body term ($t_3$).

A widely-used variant of the Skyrme interaction replaces the three-body interaction with a two-body density-dependent form \cite{Erler2010},
\begin{equation}
  \upsilon_3(\pmb{r}_1,\pmb{r}_2) = \frac{t_3}{6}\delta(\pmb{r}_1-\pmb{r}_2)\rho^\alpha(\pmb{r}_1)(1+x_3\P^\sigma),
\end{equation}
which adds a new exchange parameter $x_3$ along with a parameter $\alpha$ which is allowed to take on non-integer values, hence breaking the link between the ``interaction'' and a force, and formally requiring and EDF picture.

From this, one derives \cite{Engel1975,Perlinska2004,Lesinski2007} a Hamiltonian density, or density functional, of 

\begin{equation} \label{eq:functional}
\begin{aligned}
\mathcal{E} = \int d^3r \mathcal{H}(\pmb{r}) &= \int d^3r\sum_{t=0,1}\Bigg\{C_t^\rho[\rho_0]\rho_t^2+C_t^s[\rho_0]\boldsymbol{s}_t^2+C_t^{\Delta\rho}\rho_t\boldsymbol{\nabla}^2\rho_t \\
&+C_t^{\nabla s}(\boldsymbol{\nabla}\cdot\boldsymbol{s})^2+C_t^{\Delta s}\boldsymbol{s}_t\cdot\boldsymbol{\nabla}^2\boldsymbol{s}_t+C_t^{\tau}(\rho_t\tau_t-\boldsymbol{j}_t^2)  \\
&+C_t^T\left(\boldsymbol{s}_t\cdot\boldsymbol{T}_t-\sum_{\mu,\nu=x}^zJ_{t,\mu\nu}J_{t,\mu\nu}\right) \\
&+C_t^F\Bigg[\boldsymbol{s}_t\cdot\boldsymbol{F}_t-\frac{1}{2}\left(\sum_{\mu=x}^zJ_{t,\mu\mu}\right)^2-\frac{1}{2}\sum_{\mu,\nu=x}^zJ_{t,\mu\nu}J_{t,\nu\mu}\Bigg]\\
&+C_t^{\nabla\cdot J}\left(\rho_t\boldsymbol{\nabla}\cdot\boldsymbol{J}_t+\boldsymbol{s}_t\cdot\boldsymbol{\nabla}\times\boldsymbol{j}_t\right)\Bigg\},
\end{aligned}
\end{equation}

Here, the summation index $t$ runs over values $0$ for isoscalar densities ($\rho_0=\rho_p+\rho_n$ and similarly for the other densities) and $1$ for isovector densities ($\rho_1 = \rho_p-\rho_n$ etc), the set $\{C_t\}$ are the coefficients of the functional and the densities are defined in terms of the density matrix
\begin{equation}
\rho_q(\boldsymbol{r}\sigma,\boldsymbol{r}'\sigma')=\frac{1}{2}\rho_q(\boldsymbol{r},\boldsymbol{r}')\delta_{\sigma\sigma'}+\frac{1}{2}\boldsymbol{s}_q(\boldsymbol{r},\boldsymbol{r}')\cdot\langle\sigma'|\hat{\boldsymbol{\sigma}}|\sigma\rangle,
\end{equation}
with the particle density matrix being
\begin{equation}
\rho_q(\boldsymbol{r},\boldsymbol{r}')=\sum_\sigma\rho_q(\boldsymbol{r}\sigma,\boldsymbol{r}'\sigma')
\end{equation}
and the spin density matrix
\begin{equation}
\boldsymbol{s}_q(\boldsymbol{r},\boldsymbol{r}')=\sum_{\sigma\sigma'}\rho_q(\boldsymbol{r}\sigma,\boldsymbol{r}'\sigma')\langle\sigma|\hat{\boldsymbol{\sigma}}|\sigma\rangle.
\end{equation}
The further densities found in the functional (\ref{eq:functional}) are given by

\begin{eqnarray}\label{eq:densities}
\rho_q(\boldsymbol{r})&=&\left.\rho_q(\boldsymbol{r},\boldsymbol{r}')\right|_{\boldsymbol{r}=\boldsymbol{r}'}\nonumber \\
\boldsymbol{s}_q(\boldsymbol{r}) &=& \left.\boldsymbol{s}_q(\boldsymbol{r},\boldsymbol{r}')\right|_{\boldsymbol{r}=\boldsymbol{r}'}\nonumber \\
\tau_q(\boldsymbol{r}) &=& \left.\boldsymbol{\nabla}\cdot\boldsymbol{\nabla}'\rho_q(\boldsymbol{r},\boldsymbol{r}')\right|_{\boldsymbol{r}=\boldsymbol{r}'} \nonumber \\
T_{q,\mu}(\boldsymbol{r}) &=& \left.\boldsymbol{\nabla}\cdot\boldsymbol{\nabla}'s_{q,\mu}(\boldsymbol{r},\boldsymbol{r}')\right|_{\boldsymbol{r}=\boldsymbol{r}'}\\
\boldsymbol{j}_q(\boldsymbol{r}) &=& \left.-\frac{i}{2}(\boldsymbol{\nabla}-\boldsymbol{\nabla}')\rho_q(\boldsymbol{r},\boldsymbol{r}')\right|_{\boldsymbol{r}=\boldsymbol{r}'}\nonumber \\
J_{q,\mu\nu}(\boldsymbol{r})&=&\left.-\frac{i}{2}(\nabla_\mu-\nabla_\mu')s_{q,\nu}(\boldsymbol{r},\boldsymbol{r}')\right|_{\boldsymbol{r}=\boldsymbol{r}'}\nonumber\\
J_{q,\kappa}(\boldsymbol{r})&=&\sum_{\mu,\nu=x}^z\epsilon_{\kappa\mu\nu}J_{q,\mu\nu}(\boldsymbol{r})\nonumber \\
F_{q,\mu}(\boldsymbol{r})&=&\left.\frac{1}{2}\sum_{\nu=x}^z(\nabla_\mu\nabla_\nu'+\nabla_\mu'\nabla_\nu)s_{q,\nu}(\boldsymbol{r},\boldsymbol{r}')\right|_{\boldsymbol{r}=\boldsymbol{r}'}\nonumber.
\end{eqnarray}
Here, as in (\ref{eq:functional}), the Greek letter indices run over the Cartesian coordinates $x, y, z$.

The densities $\rho$, $\tau$, and $\mathbb{J}$ are time-even\footnote{We write $\mathbb{J}$ without the coordinate subscript indices for the tensor quantity, and $J_{q,\mu\nu}$ for the scalar quantity that comes from specifying subscripts} (identical upon reversal of the sign of the time coordinate), while $\boldsymbol{s}$, $\boldsymbol{T}$, and $\boldsymbol{F}$ are time-odd (change sign upon change of the sign of $t$).  Terms in the Hamiltonian density (\ref{eq:functional}) are all time-even, and are made of bilinear products of either two time-even densities or two time-odd densities \cite{Dob-Dud}.  Time-odd densities are identically zero in the ground states of even-even nuclei and are essentially unconstrained by fits of the Skyrme interaction parameters, which are made to ground states of even-even nuclei and to nuclear matter properties.  While all the terms in the Hamiltonian are time-even, the phrase ``time-odd terms'' is used to mean those terms made of time-odd densities.

In making the derivation from the interaction to the functional, there is a fixed link between the two sets of coefficients \cite{Engel1975,Barton2018}.  One can choose to either break this link or not, and to fit either set of parameters directly.  So far, the majority of fitted sets of parameters in the literature \cite{Dutra2012} keep the link and fit at the level of the interaction parameters.  Note that the terms in the functional (\ref{eq:functional}) which feature derivatives of the spin density apparently give rise to instabilities \cite{Kortelainen2010,Fracasso2012} and are not usually included in actual calculations.

If linking the interaction parameters to the density functional coefficients, one has the choice of using only those terms in the density functional which are really constrained at the fitting stage -- i.e. those that are associated with non-zero terms in the ground states of even-even nuclei or nuclear matter (or indeed, the subset of these terms which were actually considered at the fitting stage) -- or one may choose to activate all terms in the functional.  Both methods are used in the literature.  Particular terms in the functional are obliged to be grouped together due to Galilean invariance.  For example, the spin-orbit interaction consists of a time-even term $C_t^{\nabla\cdot J}\rho_t\boldsymbol{\nabla}\cdot\boldsymbol{J}_t$ and a time-odd term $C_t^{\nabla\cdot J}\boldsymbol{\nabla}\times\boldsymbol{j}_t$ with the same coefficient.  If these two terms are allowed to have different coefficients, then Galilean invariance is broken and, for example, a calculation translating a nucleus through space will fail to conserve energy \cite{Maruhn2006}.

Since the focus of this review is on the effect of the interactions, we give here coefficients of the functional in terms of those of the interaction, so that one may clearly see from which terms in the interaction (\ref{eq:twobodyforce}) the terms in the functional (\ref{eq:functional}) arise:
\bea
C_0^\rho &=& \frac{3}{8}t_0 + \frac{3}{48}t_3\rho_0^\alpha(\pmb{r}),\\
C_1^\rho &=& -\frac{1}{4}t_0(\frac{1}{2}+x_0)-\frac{1}{24}t_3(\frac{1}{2}+x_3)\rho_0^\alpha(\pmb{r}),\\
C_0^s &=&-\frac{1}{4}t_0(\frac{1}{2}-x_0)-\frac{1}{24}t_3(\frac{1}{2}-x_3)\rho_0^\alpha(\pmb{r}),\\
C_1^s &=& -\frac{1}{8}t_0 -\frac{1}{48}t_3\rho_0^\alpha(\pmb{r}),\\
C_0^\tau &=& \frac{3}{16}t_1 + \frac{1}{4}t_2(\frac{5}{4}+x_2),\\
C_1^\tau &=& -\frac{1}{8}t_1(\frac{1}{2}+x_1)+\frac{1}{8}t_2(\frac{1}{2}+x_2),\\
C_0^T &=& -\frac{1}{8}t_1(\frac{1}{2}-x_1)+\frac{1}{8}t_2(\frac{1}{2}+x_2)-\frac{1}{8}(t_e+3t_o),\\
C_1^T &=& -\frac{1}{16}(t_1-t_2)-\frac{1}{8}(t_e-t_o),\\
C_0^{\Delta\rho} &=& -\frac{9}{64}t_1 + \frac{1}{16}t_2(\frac{5}{4}+x_2),\\
C_1^{\Delta\rho} &=& \frac{3}{32}t_1(\frac{1}{2}+x_1)+\frac{1}{32}t_2(\frac{1}{2}+x_2),\\
C_0^{\Delta s} &=& \frac{3}{32}t_1(\frac{1}{2}-x_1) + \frac{1}{32}t_2(\frac{1}{2}+x_1) - \frac{3}{32}(t_e-t_o),\\
C_1^{\Delta s} &=& \frac{1}{64}(3t_1 + t_2) - \frac{1}{32}(3t_e+t_o),\\
C_0^{\nabla s} &=& -\frac{9}{32}(t_e-t_o),\\
C_1^{\nabla s} &=& -\frac{3}{32}(3t_e+t_o),\\
C_0^{\nabla J} &=& -\frac{3}{4}W_0,\\
C_1^{\nabla J} &=& -\frac{1}{4}W_0.
\eea

From the energy density, one attempts to find the optimal solution by varying with respect to each of the densities:

\begin{equation}
\begin{aligned}
\delta \mathcal{E} \ = \ \sum_{q} \int d^{3} r \bigg\{ \dfrac{\partial H}{\partial \tau_{q}} \delta \tau_{q} + \dfrac{\partial H}{\partial \rho_{q}} \delta \rho_{q}  + \sum_{\mu \nu} ( \dfrac{\partial H}{\partial J_{q, \mu \nu}} \delta J_{q, \mu \nu} ) + \sum_{\mu} \Big( \dfrac{\partial H}{\partial J_{q,\mu} } \delta J_{q,\mu} \\ + \dfrac{\partial H}{\partial j_{q,\mu}}  \delta j_{q,\mu} + \dfrac{\partial H}{\partial T_{q,\mu}} \delta T_{q,\mu} + \dfrac{\partial H}{\partial s_{q,\mu}} \delta s_{q,\mu}  + \dfrac{\partial H}{\partial F_{q,\mu}}  \delta F_{q,\mu} \Big) \bigg\} .
\end{aligned}
\end{equation}
Here, we have switched to a form in which neutron and proton densities (labelled by $q$) are treated separately, rather than as isoscalar and isovector sums and differences.  This reflects the usual computational implementation strategy.

The partial derivatives are conventionally written in symbolic form as
\be
\begin{aligned}  \label{deltaE}
\delta E \ &= \ \sum_{q} \int d^{3} r \bigg\{ \dfrac{\hbar^{2}}{2m_{q}^{*}} \delta \tau_{q} + U_{q} \delta \rho_{q} + \sum_{\mu \nu} ( \gamma_{q, \mu \nu} \delta J_{q, \mu \nu} ) + \sum_{\mu} \Big( B_{q,\mu} \delta J_{q,\mu} \nonumber \\
&+ I_{q,\mu} \delta j_{q,\mu} + C_{q,\mu} \delta T_{q,\mu} + \Sigma_{q,\mu}  \delta s_{q,\mu}  + D_{q,\mu} \delta F_{q,\mu} \Big) \bigg\}.
\end{aligned}
\ee

Since the densities are made up of single-particle wave functions, the variation of each kind of density amounts to the variation of the single particle wave functions.  Combining a minimisation of the energy along with a Lagrange multiplier constraint to ensure normality of each single particle wave function,

\begin{equation} \label{constraint}
\delta  \bigg( E - \sum_{\alpha=1}^{A} e_{\alpha} \int d^{3} r \phi_{\alpha}^{*} (\pmb{r}) \phi_{\alpha} (\pmb{r})  \bigg) \ = \ 0,
\end{equation}

one arrives at the Kohn-Sham equations which represent the particular method of approaching DFT in which one considers the density to comprise single particle wave functions:

\be
\begin{aligned} \label{eq:ksham}
&\Bigg[ - \pmb{\nabla} \cdot \bigg( \dfrac{\hbar^{2}}{2m_{q}^{*} (r)} \pmb{\nabla} \bigg) + U_{q} + \dfrac{1}{2i} \sum_{\substack{\mu \nu \\ \sigma \sigma'}} \bigg( (\pmb{\nabla} \cdot \pmb{\sigma}) \gamma_{q, \mu \nu} + \gamma_{q, \mu \nu}  (\pmb{\nabla} \cdot \pmb{\sigma})  \bigg) +  \\
&\dfrac{1}{i} \pmb{B}_{q} \cdot \Big(\pmb{\nabla} \times \pmb{\sigma} \Big) - \pmb{\nabla} \cdot \bigg( (\pmb{\sigma} \cdot \pmb{C}_{q}) \pmb{\nabla} \bigg) + \pmb{\sigma} \cdot \pmb{\Sigma}_{q} + \dfrac{1}{2i} \bigg( \pmb{\nabla} \cdot \pmb{I}_{q} + \pmb{I}_{q} \cdot \pmb{\nabla} \bigg)  \\ 
&- \dfrac{1}{2} \sum_{\substack{\mu \nu \\ \sigma \sigma'}} \pmb{\sigma}_{\nu, \sigma \sigma'} \bigg( (\nabla_{\nu} D_{q, \mu}) \nabla_{\mu} + 2 D_{q, \mu} \nabla_{\nu} \nabla_{\mu}  + (\nabla_{\mu} D_{q, \mu}) \nabla_{\nu} \bigg) \Bigg] \phi_{\alpha} \ = \ e_{\alpha} \phi_{\alpha},
\end{aligned}
\ee
where the quantities in these terms are given in terms of Skyrme force parameters in Appendix \ref{sec:ksterms}.

The quantity between large square brackets acting on the left hand side on the single particle wave function is thus identified with the single particle Hamiltonian as used in the HF and TDHF equations.

This is a complete specification of the Skyrme-Kohn-Sham Hamiltonian making no assumptions for symmetries and including the tensor terms in the Skyrme interaction.  Actual Skyrme parameter sets used in the literature may have been fitted using a subset of this full Hamiltonian, and one should be aware of the detailed form of the interaction used when fitting a parameter set before making use of it oneself.  A derivation of the Hamiltonian assuming time-reversal and axial symmetry was originally given by Vautherin and Brink \cite{Vautherin1972}.  Engel et al. \cite{Engel1975} extended the derivation to allow time-reversal symmetry breaking, as necessary for any dynamic calculation and for triaxial and odd-mass static calculations.  Their version of the Skyrme interaction assumed $x_1=x_2=t_o=t_e=0$.  A complete specification of the mean-field without detailed derivation was given by Perli\'nska et al. \cite{Perlinska2004}.  Full derivations of the expressions given in Appendix \ref{sec:ksterms} are available in unpublished theses \cite{Flocard1975,Barton2018}, the most recent of which, while unpublished, is freely available from the awarding institute's online repository.

\subsection{Pairing}
The pairing interaction is important in the determination of ground-state properties of open-shell nuclei.  Its role in most aspects of heavy-ion reaction dynamics is thought to be relatively unimportant, however, its role has been studied in heavy-ion collisions \cite{Ebata2013} and has been shown to be significant in transfer reactions \cite{Scamps2013} and in other large-amplitude collective motion, such as fission \cite{Bulgac2016}.  Systematic studies of the variation of the effective pairing interaction on the behaviour of heavy-ion dynamics has not been extensively studied.  

\subsection{The BKN interaction}
In early TDHF calculations, a simplified version of the Skyrme interaction, which became known as the BKN interaction was used \cite{Bonche1976}.  It takes the $t_0$ and $t_3$ terms of the original Skyrme interaction (\ref{eq:twobodyforce}) and replaces the momentum-dependent terms with a finite-range Yukawa potential with exchange coefficients constructed to yield an action in the mean field solely in the direct term.  This results in an energy density functional of \cite{Bonche1976}
\begin{equation}
  \mathcal{H}(\pmb{r}) = a_0\tau(\pmb{r}) + \frac{3}{4}t_0\rho(\pmb{r}) + \frac{3}{16}t_3\rho^2(\pmb{r}) + V_0\int\rho(\mathbf{r}')\frac{e^{-|\mathbf{r}-\mathbf{r}')|/a}}{|\mathbf{r}-\mathbf{r}'|/a} d\mathbf{r}'
  \end{equation}
Note that the spin-orbit interaction is specifically not included in the BKN force.

\section{Comparison of effective interactions in TDHF} \label{sec:tdhfresults}
\subsection{Early TDHF calculations}

\begin{figure}[tb]
\begin{center}
\begin{minipage}[t]{10 cm}
\includegraphics[width=10cm]{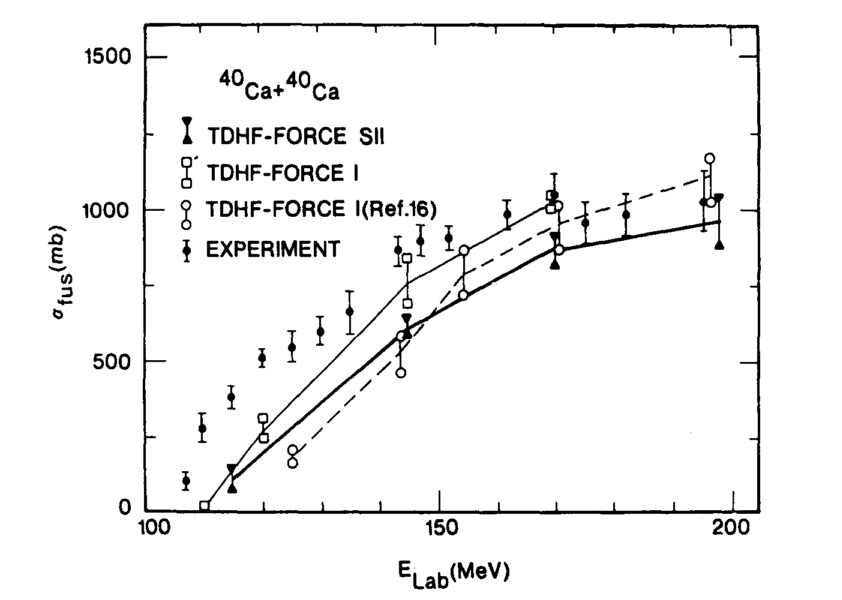}
\end{minipage}
\begin{minipage}[t]{16.5 cm}
  \caption{Early TDHF results, from \cite{Koonin1980}, in which the results of different Skyrme forces are compared.  ''Force I (Ref. 16)'' is the BKN force \cite{Bonche1976}, ''Force I'' is the BKN force with relaxed isospin symmetry and ``Force SII'' is the SII full Skyrme interaction \cite{Vautherin1972}.  Note that Ref 16. of \cite{Koonin1980} is \cite{Davies1978}.
    \label{fig:ca40ca40early}}
\end{minipage}
\end{center}
\end{figure}

The first nuclear TDHF calculations were made in the 1970s \cite{Koonin1976,Bonche1976,Cusson1976,Koonin1980}, featuring simplified versions of the Skyrme interaction, and/or restricted geometries. Figure \ref{fig:ca40ca40early}, from an early review paper \cite{Koonin1980}, shows a comparison between the experimental fusion cross-section for $^{40}$Ca + $^{40}$Ca collisions compared with two different implementations of the BKN force, and the SII \cite{Vautherin1972} Skyrme interaction.  One sees that there are noticeable effects in the calculated cross sections both from the specific choice of force parameters, as well as the allowed symmetries underlying the implementation.  By relaxing the isospin symmetry with the BKN force, the cross section increases, thanks to the ability for the initial translational kinetic energy to transfer into internal collective excitation modes permitted through the relaxation of symmetry.

The use of the BKN force vs the full Skyrme force was motivated by relative ease of implementation, though the genuine finite range of the Yukawa terms may be considered more physical than the zero-range momentum-dependent terms.  As well as omitting those terms from the energy density functional whose coefficients feature the $t_1$ and $t_2$ terms, the lack of the momentum-dependent terms gave a fixed effective mass of $m^*/m=1$.

An early study including the numerically complicated momentum-dependent terms brought them in at the level of the density-dependent effective mass \cite{Dhar1979}.  Here, versions of Skyrme forces SII \cite{Vautherin1972}, SIII, SIV, SV, and SVI \cite{Beiner1975} in which Yukawa terms are used in place of the momentum-dependent terms, except for the effective mass (i.e., (\ref{eq:effmass}) is implemented in full assuming $t_1\neq0$ and $t_2\neq0$, but $t_1=t_2=0$ elsewhere in the Skyrme mean field).  These Yukawa versions of the Skyrme forces are re-fitted to agree with the original Skyrme forces in nuclear matter.  The authors of this study found that in head-on collisions of $^{16}$O+$^{16}$O the upper fusion threshold was strongly dependent on $m^*/m$, which has a strong influence on the time-scale of the first reflection of single particle wave functions from the potential wall following collisions.  The reflected wave functions then 're-flood' the neck thus acting against the separation of the two fragments.

\subsection{Spin-orbit interactions}
The earliest Skyrme-like TDHF calculations did not include the spin-orbit interaction, owing to the complication of its implementation and the desire to at least make calculations of e.g. spin-orbit-saturated $^{16}$O collisions without the spin-orbit force to learn the first results from semi-realistic TDHF calculations.

The first implementation of the Skyrme interaction's spin-orbit force came in the mid-1980s by Umar, Strayer and Reinhard \cite{Umar1986}, with further elaboration coming from these authors plus collaborators \cite{Reinhard1988,Umar1989}.  Inclusion of the spin-orbit interaction has a dramatic effect of the dynamics of heavy-ion reactions, since it couples together the spatial motion of the nucleons with the spin degree of freedom, and gives a mechanism for kinetic energy of the incoming nuclei to strongly excite internal spin degrees of freedom.  The spin-orbit force is responsible for resolving the so-called ``fusion window anomaly'' which was found in the earliest calculations, whereby TDHF calculations gave conspicuous transparency for central collisions.  Such transparency was not observed despite extensive searches motivated by the theoretical results \cite{Kox1980,Lazzarini1981,DeToledo1981,Ikezoe1986,Auger1988}.

\begin{figure}[tb]
\begin{center}
\begin{minipage}[t]{10 cm}
\includegraphics[width=10cm]{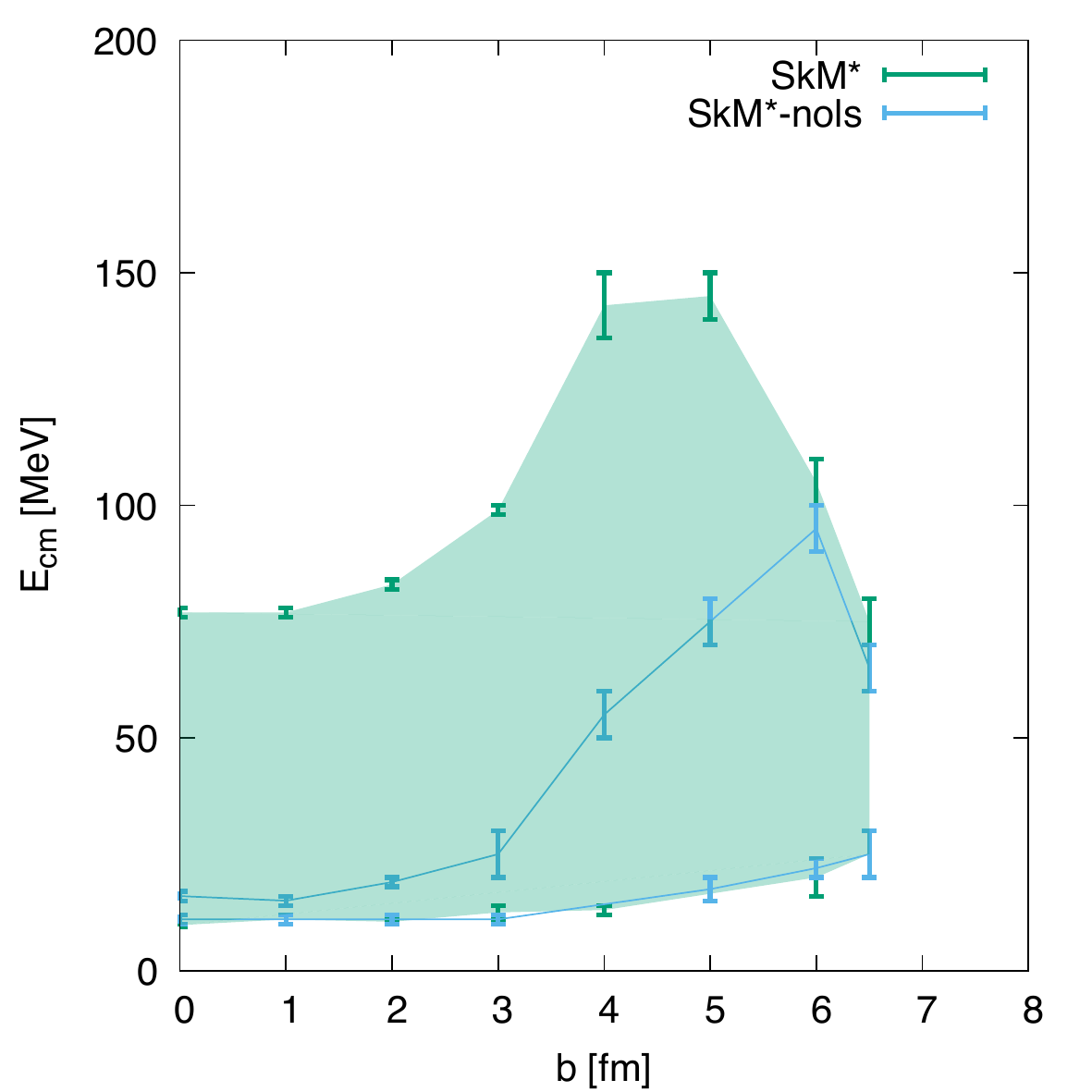}
\end{minipage}
\begin{minipage}[t]{16.5 cm}
  \caption{Fusion landscape for SkM*, as fig. \ref{fig:window_skms}, but with the much-reduced fusion region indicated for the case of an absent spin-orbit interaction with the SkM* interaction (i.e. a modified SkM* interaction in which $W_0=0$). \label{fig:skmslswindow}}
\end{minipage}
\end{center}
\end{figure}

Figure \ref{fig:skmslswindow} shows the fusion landscape in the $E_{\mathrm{cm}}$--$b$ plane for the SkM* force both with and without the spin-orbit force.  The shaded region shows the locus of fusion for the full SkM* interaction, while the lines indicated by ``SkM*-nols'' show the smaller region for fusion when the spin-orbit force is absent.  One notices at small impact parameter that fusion occurs in the absence of the spin-orbit interaction only over a very limited range of energies.  For the most peripheral reactions that result in fusion -- i.e. for large impact parameter $b$ -- the effect of the spin-orbit force is much diminished.  This is because very little kinetic energy is being turned into internal inelastic excitation, but the capture and fusion depends more upon the tail of the densities being able to form a neck to form a rotating compound nucleus with relatively little internal spin excitation.  The striking increase observed with the spin-orbit interaction for small $b$ effectively resolved the fusion window anomaly.  The original work \cite{Umar1986} examined the dependence upon the Skyrme interaction by using forces SII \cite{Vautherin1972} and SkM* \cite{Bartel1982} both with and without spin-orbit, and found similarly large significant effects in both cases.  

The Fusion window anomaly has subsequently been revisited within the TDHF picture to assess the extent to which the TDHF approximation itself, with its restriction to one-body dynamics, might be responsible for unwonted transparency.  Tohyama and Umar \cite{Tohyama2002} used an extended form of TDHF, known as TDDM \cite{Tohyama1987,Gong1990,Gong1990a} in which certain aspects of the dynamics of the two-body density matrix, and explicit two-body collisions, are taken into account.  They found that the extra dissipation allowed by the TDDM approximation was almost as significant as the spin-orbit interaction:  The upper fusion threshold increased from 30\ MeV to 69\ MeV due to the spin-orbit interaction and from 30\ MeV to 66\ MeV due to two-body collisions (but in the absence of spin-orbit).  with both effects, the increase is to 80\ MeV.

In table \ref{tab:fust} details of all known TDHF calculations which map out the upper fusion limit for $^{16}$O+$^{16}$O at zero impact parameter are presented, including those cases where the spin-orbit force has been deliberately switched on or off, and including TDDM results.

\begin{figure}[tb]
\begin{center}
\begin{minipage}[t]{10 cm}
\includegraphics[width=10cm]{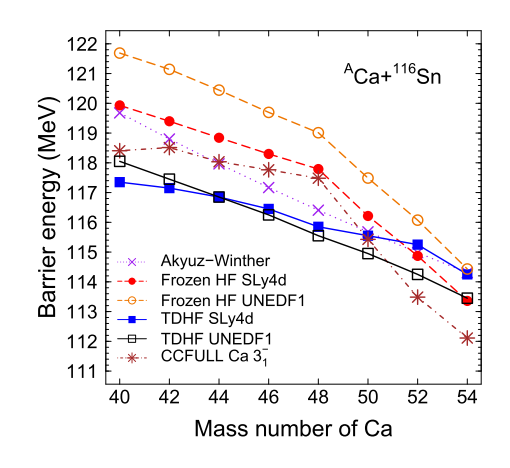}
\end{minipage}
\begin{minipage}[t]{16.5 cm}
  \caption{Barrier energies $^{A}$Ca + $^{116}$Sn reactions as a function of mass number of the calcium nucleus.  From \cite{Vo-Phuoc2017} \label{fig:casn}}
\end{minipage}
\end{center}
\end{figure}

The standard form of the spin-orbit potential from the Skyrme interaction's spin-orbit potential is (see (\ref{eq:spino}))
\be
\pmb{B}_{q} \ = \ \dfrac{W_{0}}{2} \pmb{\nabla} (\rho + \rho_{q}).
\ee
This one-parameter form has a fixed isospin dependence and its posited form is motivated in part through its simplicity.  In relativistic mean field (RMF) approaches \cite{Reinhard1989}, in which the spin-orbit term arises naturally, the spin-orbit potential's isospin dependence comes in a form proportional to $\pmb{\nabla}\rho$ rather than Skyrme's $\pmb{\nabla} (\rho + \rho_{q})$ while the strength has a density dependence.  Various extensions to the Skyrme mean field have been proposed to explore more general spin-orbit forces \cite{Pearson1994,Onsi1997,Xu2015}, motivated by the Relativistic Mean Field.  The simplest extension comes from allowing one extra parameter vary the isospin dependence as \cite{Reinhard1995}
\be \label{eq:bform}
\pmb{B}_{q} \ = \ b_4 \pmb{\nabla}\rho + b'_4\pmb{\nabla}\rho_{q}.
\ee
If $b'_4/b_4 = 1$ the Skyrme mean field is recovered, while $b'_4/b_4 = 0$ gives the relativistic mean field.  For a standard Skyrme force, $b_4=b'_4=\frac{W_0}{2}$.  Of course, other choices of $b'_4/b_4$ are possible, and parameter sets have been developed with this generalised spin-orbit form which have then been used in TDHF calculations. The SkIx sets, from the original paper by Reinhard and Flocard \cite{Reinhard1995} have been used for the study of fusion barriers \cite{Vo-Phuoc2017}, ternary fusion \cite{Iwata2013}, in the study of equilibration within TDHF \cite{Loebl2011}, and giant resonance calculations \cite{Goddard2013,Stevenson2015b}.  Of relevance for the case of heavy-ion reactions, Vo-Phuoc et al. \cite{Vo-Phuoc2017}, compared the barrier energy as calculated with TDHF for the SLy4d \cite{Kim1997} and UNEDF1 \cite{Kortelainen2012} Skyrme interactions.  These were chosen for comparison since they both treat the centre-of-mass correction in the same way (in that no correction is included at the Hartree-Fock level, in the spirit that and EDF should be capable of including such correlations in the fit), but differ in the form of spin-orbit interaction.  The dependence of the fusion barrier energy between the two Skyrme interactions is reproduced in figure \ref{fig:casn}.  The authors calculate the barrier energy in the Frozen HF approximation \cite{Washiyama2008,Simenel2008a} and in full TDHF.  One sees the systematic difference between the two interactions in the Frozen HF approximation, and a reduction in this difference in full TDHF dynamics.  The kink in the barrier energy at the N=28 magic number is observable in the frozen HF densities, but washed out in TDHF, presumable due to the deformation induced in the dynamics which allow orbitals either side of the magic number to be explored.  For most values of A the TDHF barrier is lower than the Frozen HF barrier.  This is to be expected since more degrees of freedom that enhance fusion open up in TDHF compared to Frozen HF.  On the other hand, for very large $A$ in $^A$Ca, the Frozen HF barrier is lower.  This is attributed to $N/Z$ equilibrium within TDHF during fragment approach, driven by the nuclear force, but increasing the Coulomb barrier.  Such equilibration is missing from the Frozen HF approach.

The SQMC parameterisation is a parameterisation of the Skyrme interaction which is fitted to reproduce as closely as possible the QMC (Quark Meson Coupling) model's mean field.  The QMC model \cite{Guichon1988,Guichon2004,Stone2016,Guichon2018} is a confined quark-level meson exchange interaction, from which a QMC EDF may be derived.  It differs slightly in functional form from the Skyrme EDF in that there are density-dependent couplings in QMC where the Skyrme EDF has point couplings, and the spin-orbit term comes out naturally from the QMC approach, with a fixed form depending on the meson couplings and masses.

\begin{figure}[tb]
\begin{center}
\begin{minipage}[t]{10 cm}
\includegraphics[width=10cm]{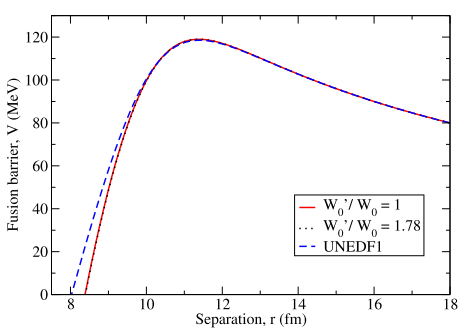}
\end{minipage}
\begin{minipage}[t]{16.5 cm}
  \caption{Frozen Hartree-Fock potentials for $^{40}$Ca + $^{132}$Sn for two forms of SQMC Skyrme functional which differ only in spin-orbit parameters, and the UNEDF1 functional. From \cite{McRae2017a} \label{fig:mcrae}}
\end{minipage}
\end{center}
\end{figure}

As a first step of exploring the QMC model in heavy-ion reactions, the Skyrme-QMC \cite{McRae2017b} parameterisation is an attempt to map the QMC energy density functional with its parameters fixed largely by the underlying quark-meson dynamics, to the Skyrme EDF.  In particular, McRae et al. \cite{McRae2017a} explored the SQMC parameter set's spin-orbit interaction properties.  In the mean-field spin-orbit potential (\ref{eq:bform}) the standard SQMC parameter sets have $b'_4/b_4=1.78$ (in contrast to the standard Skyrme value of $1.0$), and a comparison is made with the UNEDF1 functional, with $b'_4/b_4=1.86$.  A plot of the frozen Hartree-Fock NN potentials for SQMC with its natural $b'_4/b_4=1.78$ dependence, SQMC with a forced $b'_4/b_4=1.0$, and UNEDF1 is reproduced in Figure \ref{fig:mcrae} for $^{40}$Ca + $^{132}$Sn.  The conclusion is that the spin-orbit dependence per se does not have a strong influence in the barrier height or location, at least as far as the frozen Hartree-Fock approximation goes.  One might suspect the details arising in the single-particle spectrum could come become more evident in the DC-TDHF method, but further studies are called for here before reaching a stronger conclusion.  Effects on radius isotope shift have already been noted for forces with the extended spin-orbit form \cite{Reinhard1995,Sharma1995,Goddard2013a}, and one can generally expect matter radii to have an effect on NN potential.  A full TDHF implementation of the QMC EDF will be necessary to fully explore its properties in heavy-ion collisions. 

\begin{figure}[tb]
\begin{center}
\begin{minipage}[t]{10 cm}
\includegraphics[width=10cm]{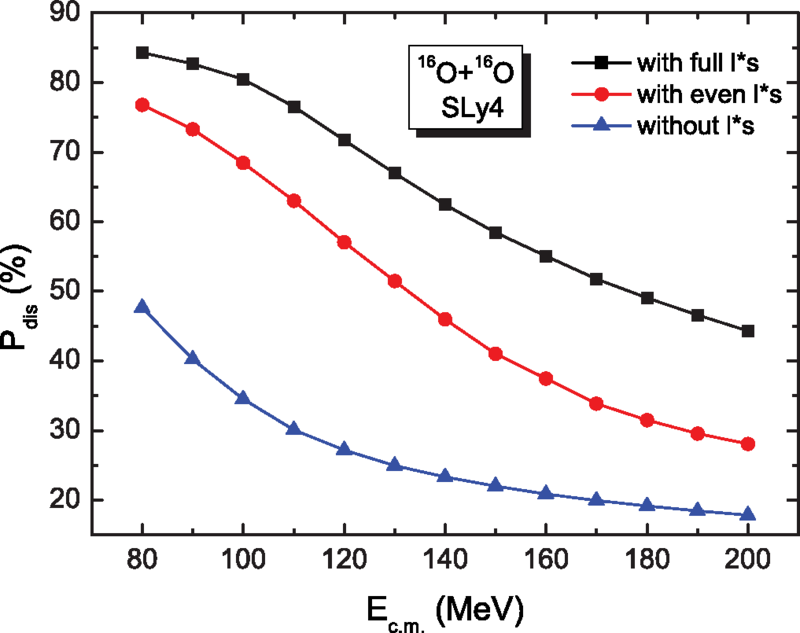}
\end{minipage}
\begin{minipage}[t]{16.5 cm}
  \caption{The percentage of relative kinetic energy which is dissipated into internal excitation of the nuclei during deep-inelastic scattering at $b=0$ as a function of initial centre of mass energy for the SLy4 Skyrme force with no spin orbit (``l*s''), with the time-even terms and with the full time-even + time-odd spin-orbit interaction. From \cite{Dai2014a}.\label{fig:pdis}}
\end{minipage}
\end{center}
\end{figure}

\begin{figure}[tb]
\begin{center}
\begin{minipage}[t]{10 cm}
\includegraphics[width=10cm]{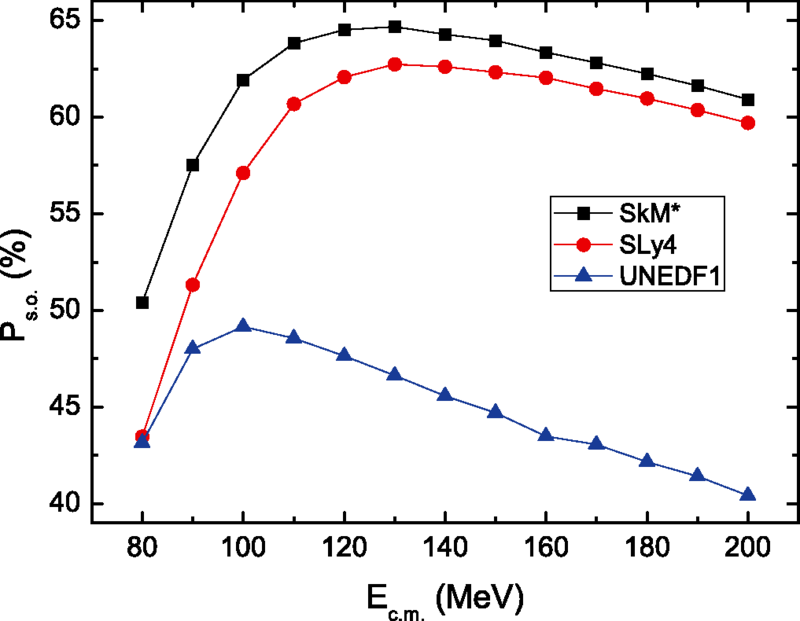}
\end{minipage}
\begin{minipage}[t]{16.5 cm}
  \caption{Proportion of energy dissipation which is due to the spin-orbit interaction in $^{16}$O + $^{16}$O deep-inelastic collisions for Skyrme forces SLy4, SkM*, and UNEDF1.  From \cite{Dai2014a}.\label{fig:pso}}
\end{minipage}
\end{center}
\end{figure}

Dai et al. \cite{Dai2014a} studied dissipation (transfer of energy from relative motion to internal excitation) effects in $^{16}$O+$^{16}$O collisions, paying particular attention to the role of the spin-orbit interaction.  They used SLy4, SkM* and UNEDF1 Skyrme interactions.  They implemented the full set of terms arising in the mean field from the spin-orbit force, including the time-even spin-orbit ($\rho\pmb{\nabla}\cdot\pmb{J}$ in the functional (\ref{eq:functional})) and the time-odd spin-orbit term ($\pmb{s}\cdot\pmb{\nabla}\times\pmb{j}$ in (\ref{eq:functional})).  They examined reactions above the upper threshold for fusion in order to explore the partial transfer of initial relative kinetic energy into a combination of final relative motion and internal excitation.  A measure of dissipation was given as
\be
P_{\mathrm{dis}}=1-E_{\mathrm{fin}}/E_{\mathrm{CM}}
\ee
where $E_{\mathrm{fin}}$ is the final relative kinetic energy between the two fragments and $E_{\mathrm{CM}}$ is the initial centre of mass energy.  When $E_{\mathrm{fin}}=0$ the nuclei are below the threshold for separation and remain fused, indicating total dissipation from collective kinetic energy to modes internal to the compound nucleus.  Figure \ref{fig:pdis} shows the enhanced dissipation caused by the inclusion of the spin-orbit interaction, as well as the increased importance of the time-odd spin-orbit force at higher initial energies (see also fig. 4 of \cite{Dai2014a}).
Following the comparison of version of the SLy4 parameter set with and without time-odd and time-even spin-orbit forces, Dai et al. go on to look at the proportion of the dissipated energy which arises from the spin orbit force, defined as
\be
P_{\mathrm{so}}=1-P_{\mathrm{dis}}^{\mathrm{(no-ls)}}/P_{\mathrm{dis}}^{(full-ls)}
\ee
where $P_{\mathrm{dis}}^{\mathrm{(no-ls)}}$ and $P_{\mathrm{dis}}^{(full-ls)}$ refer to the blue lines with triangular points and the black line with square points in figure \ref{fig:pdis} respectively.  This proportion of dissipated energy due to the spin-orbit force is shown in figure \ref{fig:pso} for SkM*, SLy4, and UNEDF1.  There is a striking difference between UNEDF1 on the one hand and SkM* and SLy4 on the other, with the UNEDF1 dissipation being much less due to its spin-orbit interaction.  It is possible that the $b'_4/b_4$ value as discussed by McRae \cite{McRae2017a}, and above, causes this remarkable effect.

In \cite{Iwata2012}, Iwata examines dissipation mechanisms by extracting a collective potential energy and following it as a function of time in collisions of $^{16}$O and $^{16}$O at 40 MeV centre-of-mass energy.  SLy4d \cite{Kim1997} and SkM* \cite{Bartel1982} Skyrme interactions are used.  The author performs calculations for each of these forces with the spin-orbit interaction turned off, as well as on, to reproduce the result that the spin-orbit interaction is crucial for lowering the fusion threshold thanks to its role in dissipation.  The results of Sly4d and SkM* are discussed as being qualitatively identical, in the sense that fusion does not occur with either force if spin-orbit is removed, but does occur when it is included.  From the provided plots, though, one sees differences of the order 30 MeV in the collective potential energy between the two effective interactions under consideration, at the point where the two fragments are touching.

\begin{table}[bht]
\centering
\begin{tabular}{ccc}\hline
Force &  Threshold (MeV) & Reference \\ \hline\hline
Skyrme II & 68 & \cite{Umar1986}\\
Skyrme II (no ls) & 31 & \cite{Umar1986} \\
Skyrme M* & 70 & \cite{Umar1986} \\
Skyrme M* (no ls) & 27 & \cite{Umar1986} \\
FY1 & 56 & \cite{Reinhard1988} \\
SLy4 & 68 & \cite{Dai2014a} \\
UNEDF1 & 76 & \cite{Dai2014a} \\
SkM$^*$ (basic) & 77 & \cite{Stevenson2015} \\ 
SkM$^*$ (inc. $J^2$) & 71 & \cite{Stevenson2015} \\ 
SkM$^*$ (full) & 73 & \cite{Stevenson2015} \\ 
SLy5 (full)  & 68  & \cite{Stevenson2015} \\ 
SLy5t  & 70  & \cite{Stevenson2015} \\ 
$T11$ & 60 & \cite{Dai2014} \\
$T{12}$ & 61  & \cite{Stevenson2015} \\ 
$T13$ & 67 & \cite{Dai2014} \\
$T{14}$ & 69  & \cite{Stevenson2015} \\ 
$T{22}$ & 64  & \cite{Stevenson2015} \\ 
$T{24}$ & 71  & \cite{Stevenson2015} \\ 
$T{26}$ & 82  &\cite{Stevenson2015} \\ 
$T31$ & 70 & \cite{Dai2014} \\
$T33$ & 77 & \cite{Dai2014} \\
$T{42}$ & 69  & \cite{Stevenson2015}\\ 
$T{44}$ & 79  & \cite{Stevenson2015}\\ 
$T{46}$ & 87  & \cite{Stevenson2015}\\
SII TDHF-nols & 30 & \cite{Tohyama2002}\\
SII TDHF+ls & 69 & \cite{Tohyama2002}\\
SII TDDM-nols & 66 & \cite{Tohyama2002}\\
SII TDDM+ls & 80 & \cite{Tohyama2002}\\
\hline\hline
\end{tabular}
\caption{Upper fusion threshold energies for the $^{16}$O + $^{16}$O collision using various parameterizations of the
Skyrme interaction.}
\label{tab:fust}
\end{table}

\subsection{Tensor interaction}
The tensor terms in the original Skyrme interaction (\ref{eq:skyv12}) were omitted in the original Hartree-Fock implementation \cite{Vautherin1972} since that work was restricted to ground states of spherical nuclei where the extra degrees of freedom allowed by the tensor terms extended only to details of spin-orbit interaction which were deemed beyond the necessity of first implementation where the basic spin-orbit force seemed to be adequate.  The effect of including the tensor interaction was first studied by its effect on single particle levels \cite{Stancu1977} in which the authors concluded that it only minor improvements to the reproduction of observed spin-orbit splittings.  The authors of this original work have since followed up with further explorations \cite{Brink2007,Brink2018}.

While the tensor terms had been occasionally included in implementations of the Skyrme interaction \cite{Liu1991,Tondeur1983}, a general renaissance in the use of tensor part of the effective interactions came from the interacting shell model \cite{Otsuka2005}.  New explorations with the Skyrme tensor force followed, and include a series of papers \cite{Lesinski2007,Bender2009,Hellemans2012} in which a selection of tensor parameterisations were introduced, each fitted with the same protocol as the SLy parameter sets \cite{Chabanat1996,Chabanat1997,Chabanat1998} each with different choices of isospin dependence given by the strength of the tensor parameters.  Col\`o et al. introduced a Skyrme-tensor parameterisation \cite{Colo2007} based on perturbatively adding the tensor terms to the SLy parameter set SLy5 \cite{Chabanat1997}.  A recent review of the tensor force in effective interactions by Sagawa and Col\`o \cite{Sagawa2014} gives further details of the use of tensor terms across many observables, as well as an historical summary of its implementation.

Inclusion of the Skyrme tensor interaction for heavy-ion collisions has been implemented with one of two philosophies.  One is to include the effects to the mean field from the tensor terms only to the spin-orbit interaction.  The argument here is that it is presumably the dominant effect of the tensor terms.  Moreover, inclusion of particular terms that arise in the energy density functional from the Skyrme interaction can be and have been treated as individual terms whose inclusion is never mandatory.  This basic inclusion of the tensor interaction gives rise to a term in the energy density for spherical even-even nuclei of \cite{Stancu1977}
\be
\Delta\mathcal{E} = \frac{1}{2}\alpha(\pmb{J}_n^2+\pmb{J}_p^2)+\beta \pmb{J}_n\cdot\pmb{J}_p.
\ee
where $\pmb{J}$ is the antisymmetrised part of the full $\mathbb{J}$ tensor, as defined in (\ref{eq:densities}).   Corresponding to this is a contribution to the spin-orbit potential (\ref{eq:spino}) of
\be
\begin{aligned}
  \Delta\pmb{B}_n &=\alpha\pmb{J}_n + \beta\pmb{J}_p, \\
  \Delta\pmb{B}_p &=\alpha\pmb{J}_p + \beta\pmb{J}_n. 
\end{aligned}
\ee
In fact, these terms functionally already exist in the spin-orbit potential as derived from the $t_1$ and $t_2$ terms of the central part of the Skyrme force, and the parameters $\alpha$ and $\beta$ are given by \cite{Suckling2011}\footnote{In the full symmetry-unrestricted version of the Skymre mean-field presented in this review, these terms arise from terms in (\ref{eq:gamma}) upon symmetry reduction.}
\be
\begin{aligned}
  \alpha=\frac{1}{8}(t_1-t_2-t_1x_1-t_2x_2)+\frac{5}{4}t_o,\\
  \beta = -\frac{1}{8}(t_1x_1+t_2x_2)+\frac{5}{8}(t_e+t_o).
\end{aligned}
\ee

Iwata and Maruhn \cite{Iwata2010b,Iwata2011} studied the effect of the tensor terms on the spin-orbit interaction specifically to understand the relative contribution of the tensor and spin-orbit terms and their role in dynamic spin polarisation.  They use a range of Skyrme parameterisations, including a series labelled SV-tls \cite{Klupfel2009} which include a parameter allowing the $t_1$ and $t_2$ contribution to the spin-orbit potential to be dialed on ($\eta_\mathrm{tls}=1$) or off ($\eta_\mathrm{tls}=0$).  All the SV- forces allow a fully free value of the spin-orbit $b'_4/b_4$ ratio when fitting.  Iwata and Maruhn defined a time-dependent ratio of the strength of the spin-orbit field as arising from the $\pmb{J}$ terms to those arising from the $\pmb{\nabla}\rho$ terms as $\pmb{W}^T_q/\pmb{W}^{LS}_q(t)$ and found the size of the ratio at an indicative time varied between $\sim 1\%$ and $\sim22\%$ depending on the interaction.  A rather strong mass-dependence was found, with the ratio increasing as mass increased, at least for the three calculated symmetric N=Z collisions $^{16}$O+$^{16}$O, $^{40}$Ca + $^{40}$Ca, and $^{56}$Ni+$^{56}$Ni.  The mass dependence for different Skyrme interactions is shown in figure \ref{fig:wratio}.  The tensor force is shown to be able to enhance or hinder the transfer of centre of mass motion into spin excitation during a heavy ion collision depending on the way in which the tensor and central parameters combine.  If $\alpha+\beta$ is negative, the dissipation into spin modes is enhanced.

\begin{figure}[tb]
\begin{center}
\begin{minipage}[t]{8 cm}
\includegraphics[width=8cm]{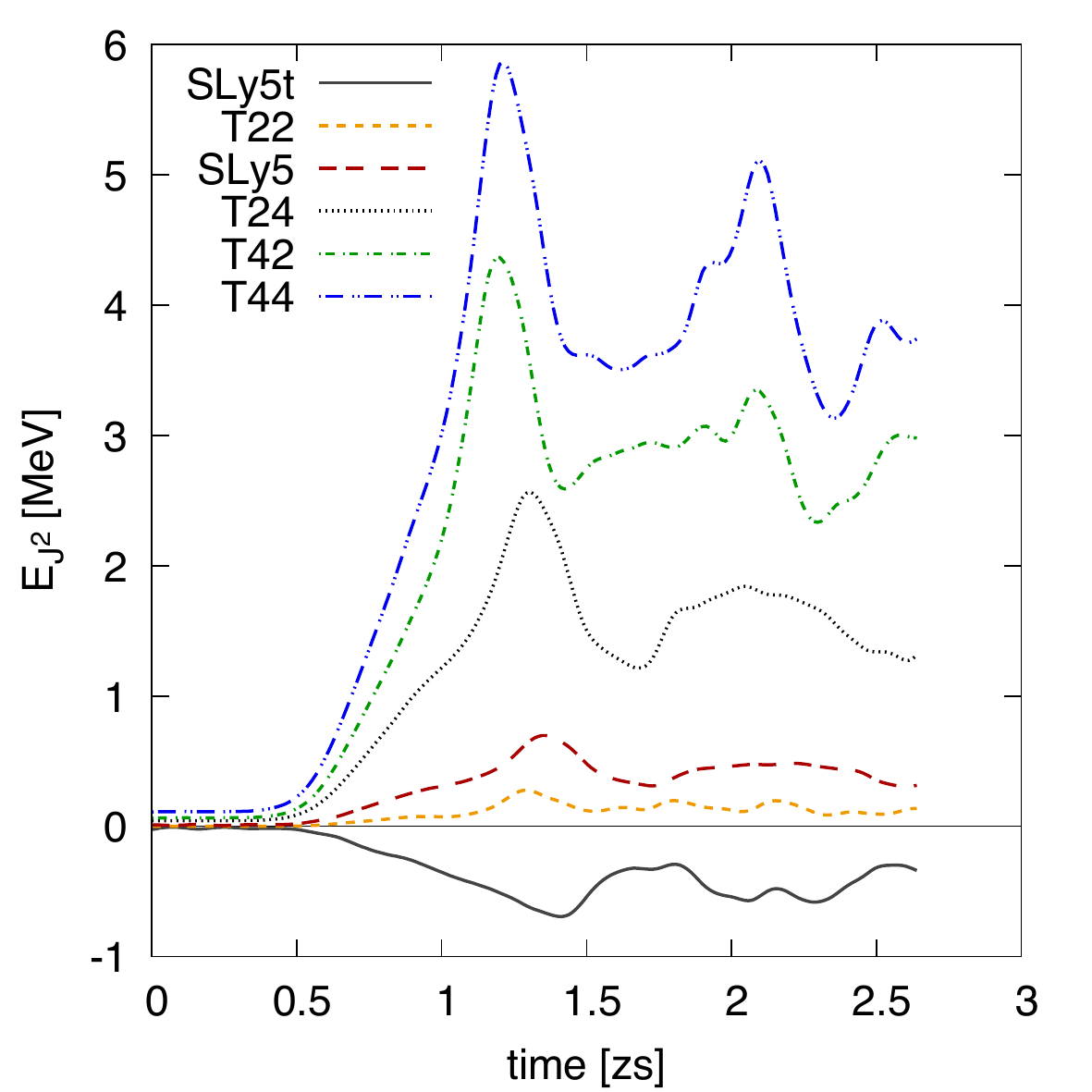}
\end{minipage}
\begin{minipage}[t]{16.5 cm}
\caption{Contribution of energies from the $\mathbb{J}^2$ terms in the Skyrme energy density functional for a selection of different Skyrme forces with tensor interaction included.  Figure adapted from \cite{Stevenson2015}.\label{j2contrib}}
\end{minipage}
\end{center}
\end{figure}

Stevenson et al. \cite{Stevenson2012,Stevenson2015,Suckling2011} used the full tensor interaction with all EDF terms except the unstable spin-dependent terms to study reactions of $^{16}$O+$^{16}$O at the upper fusion threshold between fusion and deep-inelastic reactions.  At $b=0$ a large variation in the upper fusion threshold was found, ranging between $61$ MeV (T12) and $87$ MeV (T46) from amongst the forces considered.  This study complemented previous studies of this benchnmark value, and a compilation of all known results is presented in Table \ref{tab:fust}.  The authors analysed the contribution to the total energy from different terms in the functional, and found typical changes in terms due to the tensor interaction to be of order of a few hundred keV.  The most pronounced changes were the $\mathbb{J}^2$ term, justifying its use as the first approximation to including only these terms when adding the tensor force.  The contribution from the $\mathbb{J}^2$ terms to the total energy can be positive (decreasing binding) or negative (increasing binding).  Figure \ref{j2contrib} shows the energy contribution from the $\mathbb{J}^2$ terms as a function of time during collisions of $^{16}$O + $^{16}$O at $34$ MeV and an impact parameter $b=6.65$ fm (to compare with the same setup used to explore terms in non-tensor-based Skyrme forces \cite{Umar2006})

\begin{figure}[tb]
\begin{center}
\begin{minipage}[t]{8 cm}
\includegraphics[width=8cm]{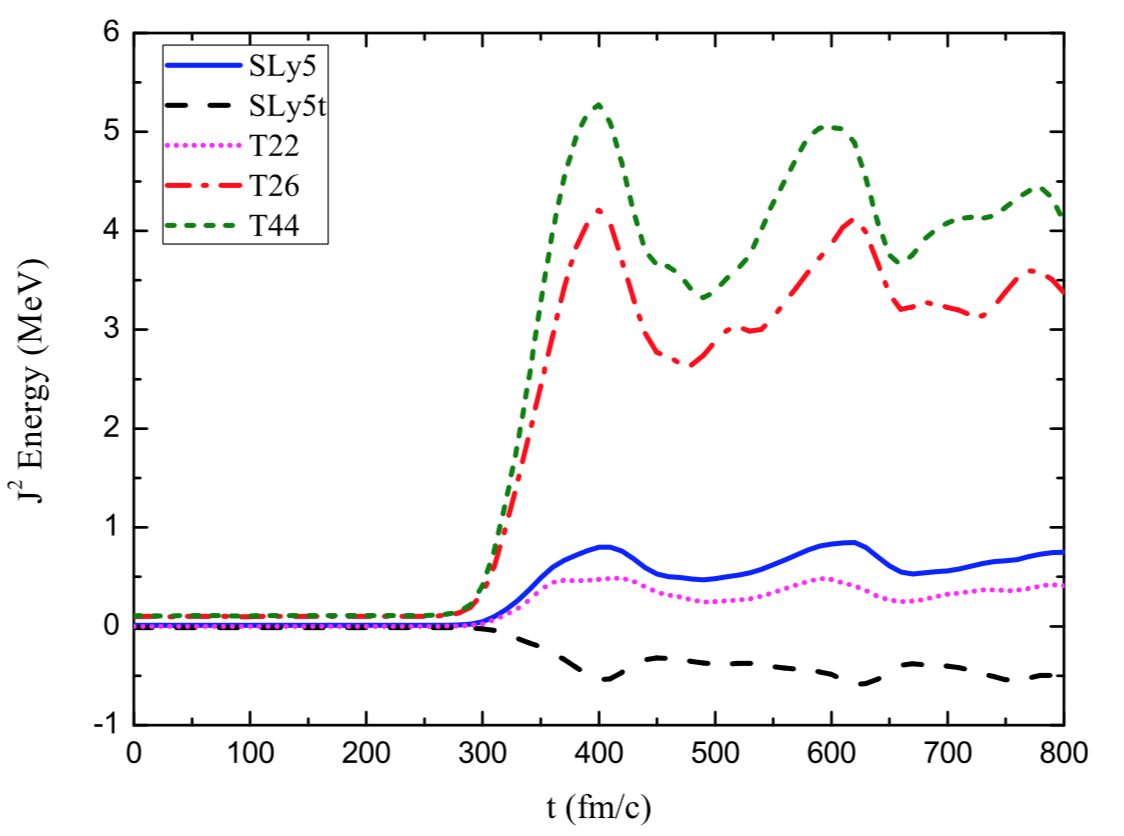}
\end{minipage}
\begin{minipage}[t]{16.5 cm}
\caption{Contribution of energies from the $\mathbb{J}^2$ terms in the Skyrme energy density functional for a selection of different Skyrme forces with tensor interaction included.  Figure from \cite{Guo2017}.\label{j2luguo}}
\end{minipage}
\end{center}
\end{figure}

\begin{figure}[tb]
\begin{center}
\begin{minipage}[t]{10 cm}
\includegraphics[width=10cm]{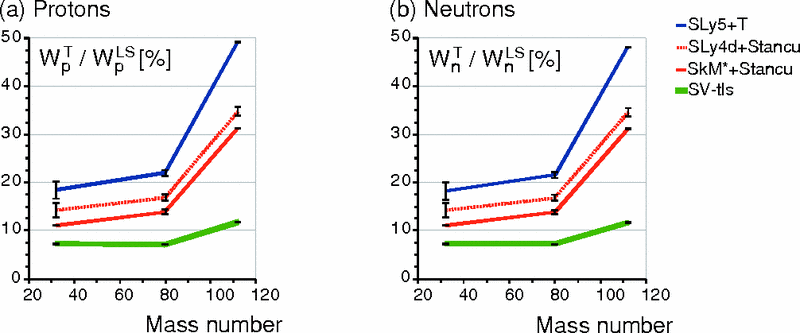}
\end{minipage}
\begin{minipage}[t]{16.5 cm}
  \caption{Relative contribution of tensor to spin-orbit force contributions to spin-orbit field as a function of mass and Skyrme interaction.  Figure from \cite{Iwata2010b}. \label{fig:wratio}}
\end{minipage}
\end{center}
\end{figure}

A similar study, extended to $^{16}$O + $^{40}$Ca \cite{Guo2017} show contributions from different parts of the energy density functional in line with that of the $^{16}$O+$^{16}$O calculations.  An analogous calculation to that of figure \ref{j2contrib} is shown in figure \ref{j2luguo}.  The same general dependence on the force parameterisation, and the possibility of increasing binding (and hence cross section) or decreasing it remains in the contribution of the $\mathbb{J}^2$ terms.

Long and Guo \cite{Long2017} use the full tensor interaction (i.e. with the full EDF, but with $A^{\nabla s}$ and $A^{\Delta s}$ again set to zero as always) and explore the fusion barrier for $^{16}$O+$^{16}$O using all 36 of the T$IJ$ tensor forces parameter sets along with $SLy5$ and $SLy5t$ within the Frozen Hartree Fock approximation.  For a selection of these forces, they calculated the barrier energy with full TDHF.  They found the height of the barrier uniformly too low.  It lay in the narrow range $9.96-10.12$ MeV, compared to an experimental value \cite{Vaz1981} of $10.61$ MeV.  The narrow range will in part be due to the spin-saturated nature of $^{16}$O and the lack of dynamical effects in the Frozen HF approximation, which render the ground state calculation quite insensitive to the tensor interaction.  Including the dynamic effects afforded with TDHF reduce the barrier heights in each case by a small amount ranging from a Frozen HF $\rightarrow$ TDHF reduction of $10.08\rightarrow10.05$ MeV (T22) to $10.02\rightarrow9.90$ (T44).  The radial separation of the nuclei at the Coulomb barrier are also systematically at variance with the data -- the Frozen HF approximation gives between $R=8.50$ fm and $R=8.58$ fm compared with an experimental value of $R=7.91$ fm.  This suggests that the tensor force does not have the right degrees of freedom to overcome any deficiencies in the ability of Skyrme-TDHF to correctly reproduce the fusion barrier in $^{16}$O+$^{16}$O.  However, this conclusion may be too hasty, and the fact that all the forces used in this study were fitted with a centre-of-mass correction, but were necessarily used without it for the two-body study.

\begin{figure}[tb]
\begin{center}
\begin{minipage}[t]{10 cm}
\includegraphics[width=10cm]{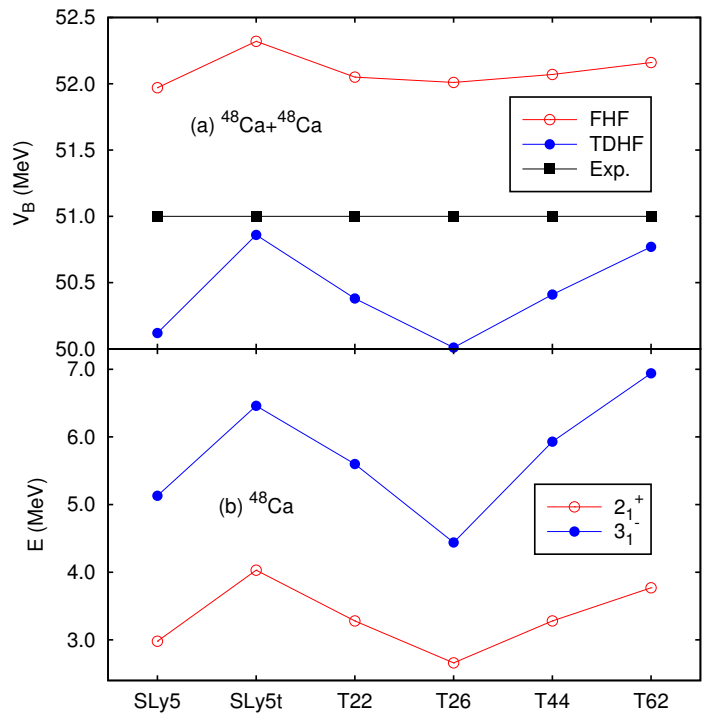}
\end{minipage}
\begin{minipage}[t]{16.5 cm}
  \caption{Upper panel shows fusion barrier for selection of Skyrme-tensor forces using the Frozen HF (FHF) approximation, or full TDHF, compared with the experimental value \cite{Stefanini2009}.  The lower panel shows collective quadrupole ($2^+$) and octupole ($3^-$) states which contribute to the lowering of the barrier in TDHF compared to FHF, with the lower energy collective states correlating with the stronger reduction in the barrier height in TDHF. Figure from \cite{Guo2018} \label{fig:ca48ca48}}
\end{minipage}
\end{center}
\end{figure}

A more positive prospect for the role of tensor parameters in fusion reactions comes from a recent study by Guo et al. \cite{Guo2018}.  They considerd reactions of $^{40}$Ca + $^{40}$Ca, $^{40}$Ca + $^{48}$Ca, $^{48}$Ca + $^{48}$Ca, $^{48}$Ca + $^{56}$Ni, and $^{56}$Ni + $^{56}$Ni, making a detailed comparison between forces for $^{48}$Ca + $^{48}$Ca as a representative example.  Figure \ref{fig:ca48ca48} shows the Frozen HF and TDHF barriers (upper panel) across the forces SLy5, SLy5t, T22, T26, T44, and T62.  As expected, the TDHF barriers are lower than those from Frozen HF.  Depending on the Skyrme paratmerisation, the barrier height can be rather well reporduced in TDHF.

\begin{figure}[tb]
\begin{center}
\begin{minipage}[t]{10 cm}
\includegraphics[width=10cm]{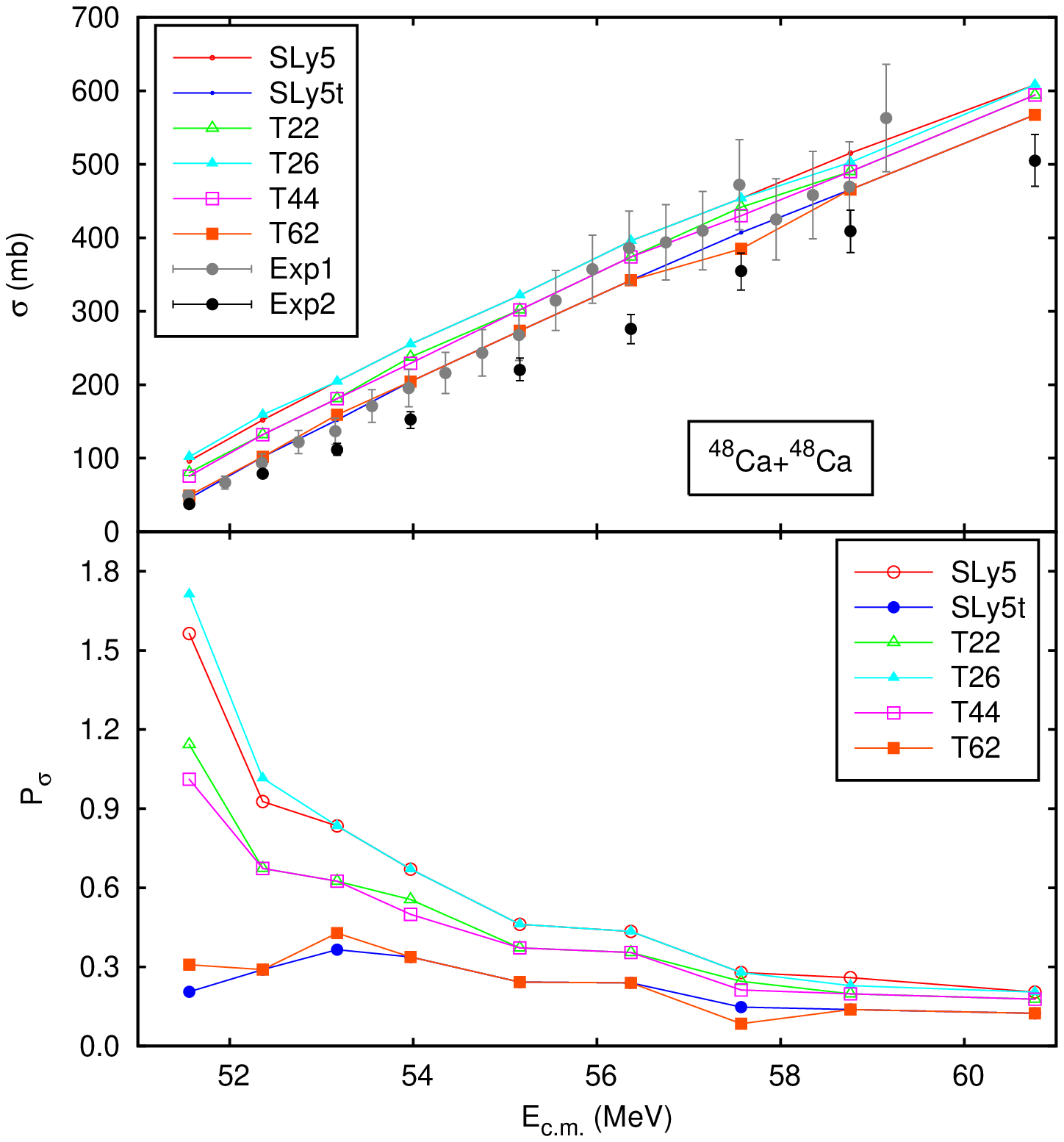}
\end{minipage}
\begin{minipage}[t]{16.5 cm}
  \caption{Cross-sections with different tensor forces, from \cite{Guo2018}. \label{fig:tensor_xsec}}
\end{minipage}
\end{center}
\end{figure}

Calculations of the cross section using the sharp-cuttoff formula (\ref{eq:xsec}) in TDHF are shown for the set of Skyrme forces considered, reproduced in figure \ref{fig:tensor_xsec}.  While all interactions overestimate the cross section, the scale of the variation between forces is such that the discrepancy between calculation and experiment, given in the lower panel of the figure and defined by
\be
P_\sigma = \frac{(\sigma_{\mathrm th}-\sigma_{\mathrm{exp}})}{\sigma_{\mathrm{exp}}}
\ee
varies markedly, and is much lower for those forces which best reproduce the barrier height in TDHF.  Thus, this work provides an example in which the Skyrme tensor force has a sufficiently large effect on reaction dynamics around the Coulomb barrier to make a changes of the order of the typical discrepancy between experimental and model calcualtions.  

Dai et al. \cite{Dai2014} studied dissipation in tensor interactions in $^{16}$O+$^{16}$O collisions in a similar manner to their work on the spin-orbit interaction \cite{Dai2014a}, looking at the energy transfer from initial relative motion to internal excitation in deep inelastic scattering.  They found that the tensor interactions could reduce dissipation compared to the SLy5 fit, or enhance it.  The T11 tensor interaction decreased the dissipation, presumably by resisting transfer of energy into the $\mathbb{J}^2$ terms, and thus has a reduced cross section compared to the other interactions studied (SLy5, SLy5+T, T11, T13, T31, T33).  The authors computed the sharp-cutoff fusion cross section in TDHF at 70.5 MeV centre of mass energy in order to compare with an experimental point of $\sigma_{\mathrm{fus}}=1056\pm125$ mb \cite{Saint-Laurent1979}.  They found that the variation between tensor parameterisations (all fitted to the same set of ground state data) differed between T11 at 1161 mb -- inside the error bar of the experimental point -- and T33 at 1327 mb.

The studies of the tensor force have therefore found that there are signigicant effects around the fusion barrier, at the top of the fusion region, and then beyond into deep-inelastic energies.

\subsection{Variation of nuclear matter properties}
\begin{figure}[tb]
\begin{center}
\begin{minipage}[t]{10 cm}
\includegraphics[width=10cm]{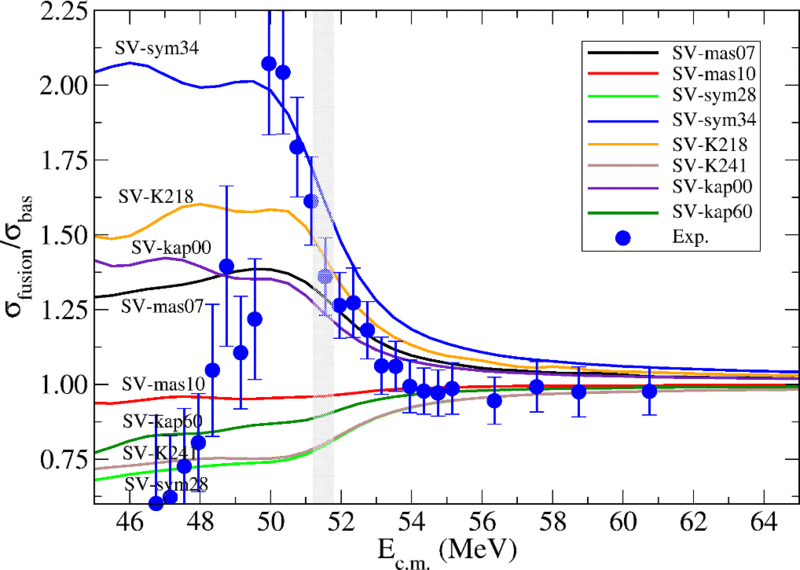}
\end{minipage}
\begin{minipage}[t]{16.5 cm}
 \caption{Cross sections for the fusion of $^{48}$Ca+$^{48}$Ca, calculated using the density-constrained TDHF method.  Shown is the ratio of the calculated cross section for a variety of different Skyrme forces from the SV- set \cite{Klupfel2009} to the basic SV parameterisation SV-bas.  Points indicate the experimental cross section with data from \cite{Stefanini2009}.  Figure from \cite{Reinhard2016}. \label{fig:nucmat}}
\end{minipage}
\end{center}
\end{figure}

The SV- range of Skyrme parameterisations \cite{Klupfel2009} were fitted each in the same manner, with each having a specific nuclear matter property (incompressibility $K$, isoscalar effective mass $m^*/m$, symmetry energy $J$, and Thomas-Reiche-Kuhn sum rule enhancement factor $\kappa_\mathrm{TRK}$ ) varied with respect to a ``basis'' parameter set SV-bas.  The SV-bas set has $K=234$ MeV, $m^*/m=0.9$, $J=30$ MeV, and $\kappa_\mathrm{TRK}=0.4$.  The parameter sets thus provide a set of interactions which can be used to study the role of nuclear matter properties in heavy-ion collisions in the Skyrme-TDHF framework.  In \cite{Reinhard2016}, the authors study fusion barriers and cross sections for $^{48}$Ca + $^{48}$Ca using a set of the SV- Skyrme interactions in order to understand the extent to which fusion cross-sections would be sensitive to nuclear matter properties.  In part this was motivated by the known link between the symmetry energy and the neutron skin thickness \cite{PhysRevLett.106.252501} that is currently a key driver of experimental determinations of neutron radii \cite{PhysRevLett.108.112502,Horowitz2014b}.  Figure \ref{fig:nucmat} shows the ratio of fusion cross section for $^{48}$Ca + $^{48}$Ca between various SV- forces, and the SV-bas base version.  The calculations are made from nucleus-nucleus potentials obtained by the DC-TDHF method (see sec. \ref{sec:dctdhf}).  The largest differences in cross section were seen by varying the symmetry energy, as expected, with the rather modest range of $J=28$--$34$ MeV (being consistent with observation \cite{Dutra2012}) spanning a range of around a factor of 3 in cross section, just below the fusion barrier.  One also sees from the figure a comparison with experimental data, showing that the modest variation of nuclear matter parameters is enough so that different model predictions differ from each other by much more than the experimental error bars (around and below the barrier, at least) and that none of the SV- forces alone fit the data across all energies.

\subsection{Other studies}

Umar and Oberacker \cite{Umar2006} made the first exploration of terms in time-odd densities that are not mandated by Galilean invariance, namely the terms in the functional (\ref{eq:functional}) in $\pmb{s}^2$, $\pmb{s}\cdot\nabla^2\pmb{s}$ and $(\pmb{s}\cdot\pmb{T}-\mathbb{J}^2)$.  Note that they did not include tensor parameterisations, so they did not need the $(\pmb{\nabla}\cdot\pmb{s})^2$ or $\pmb{s}\cdot\pmb{F}$ terms.  They observed noticeable effects in the position of the upper threshold between fusion and deep inelastic scattering when activating the time-odd terms, highlighting the need to at least consider them for inclusion in one's density functional.

In \cite{Iwata2015}, versions of the SV-bas force \cite{Klupfel2009} are generated in which whole terms in the interactions are turned on or off, to examine the resulting transparency during collisions -- and the corresponding connection to considering the travelling quantum mechanical wave packet resprsenting the nuclei to be solitons \cite{Raha1973,Foster2015,Foster2015a}.  The soliton-like nature of the wave-packet is confirmed by a high-degree of transparency when (and if) the colliding nuclei pass through each other without change.  In collisions of $^{4}$He on $^{8}$He, the transparency was found to be highly energy-dependent, with a value of around 30MeV incident energy giving high-transparency and hence soliton-like behaviour.  In terms of the interactions, it is the momentum-dependent terms in the Skyrme interaction (the $t_1$, $t_2$ and spin-orbit terms ) that suppress transparancy the most.

Loebl et al. followed up a study of dissipation in Skyrme-TDHF using the Wigner transformation \cite{Loebl2011} which concluded that full equilibration does not occur in TDHF, with a further study to see if there is any dependence on the parameterisation choice, or in the use of time-odd terms not usually activated \cite{Loebl2012}.  In particular, they performed calculations with the SLy4 \cite{Chabanat1996} force in its standard form, and also with the $\pmb{s}^2$ and $\pmb{s}\cdot{\pmb{T}}-\mathbb{J}^2$ terms from the functional (\ref{eq:functional}).  While no difference in the equilibration was found, the details of long-time differences in outcome near the upper fusion threshold were observed, with the location of the threshold being sensitive to the change in dissipation coming from the extra terms.

Godbey, Umar, and Simenel \cite{Godbey2017} took a single Skyrme interaction (SLy4) and separately calcualted contributions from the isovector and isoscalar terms in the EDF (i.e. the terms t=0 and 1 respectively in the sum in equation (\ref{eq:functional}) as applied to the DC-TDHF potential.  So doing, they were able to quantify the isovector contribution to the ion-ion potential.  For fusion reactions in which transfer channels are active, the authors showed that an isovector reduction in the potential existed, demonstrating a fusion enhancement due to transfer.  In principle one should then expect these calculated results to depend upon the isospin nature of a particualr Skyrme parameterisation.

\section{Conclusion}
The role of the effective interaction in the dynamics of heavy-ion reactions has been surveyed.  Within mean-field dynamics, the effects of varying the effective interaction between reasonable limits (i.e., using only those interactions which are available in the literature and that fit ground state data well) produces qualitatively and quantitatively variable behaviour in heavy-ion collisions at energies below the Coulomb barrier, in the fusion region, and in the deep-inelastic region at the upper energy limits where one supposes mean-field dynamics to be a reasonable approximation.  One concludes, therefore, that the role of the effective interaction in the calculation of reaction dynamics is instrumental in understanding the details of the reaction, and that results from heavy-ion reactions inform us about the details of the effective interaction.  In the case of the Skyrme-tensor interaction, both structure of individual nuclei and their dynamics as they collide can be affected.  Further study is needed on the interplay between these two aspects.

\section*{Acknowledgements}
Grateful acknowledgements are made to those with whom the authors have directly collaborted on TDHF matters: J. A. Maruhn, A. S. Umar, P.-G. Reinhard, S. Fracasso, D. Almehed, E. B. Suckling, P. M. Goddard, J. M. Broomfield, and M. R. Strayer.  Thanks are also extended to those experts with whom we have had discussion on TDHF: Ph. Chomaz, C. Simenel, L. Guo, and T. Nakatsukasa.

This work has been supported by grants from the UK STFC under grants ST/P005314/1 and ST/N002636/1, and awarded time on the STFC DiRAC computer facility.
\appendix
\addappheadtotoc
\renewcommand{\appendixname}{}
\section{Terms in the Skyrme-Kohn-Sham equation}\label{sec:ksterms}
The terms featured in the Skyrme-Kohn-Sham equations (\ref{eq:ksham}) are given below \cite{Barton2018}
\bea
\label{eq:effmass} \dfrac{\hbar^{2}}{2m_{q}^{*}} \ &=& \ \dfrac{\hbar^{2}}{2m} + \dfrac{1}{4} \Big( t_{1} + t_{2} + \dfrac{t_{1}x_{1} + t_{2}x_{2}}{2} \Big) \rho + \dfrac{1}{8} \Big( t_{2} - t_{1} +2t_{2}x_{2} -2t_{1}x_{1} \Big) \rho_{q} \\
U_{q} (r) \ &=& \ t_{0} \bigg[ \Big(1 + \dfrac{x_{0}}{2} \Big) \rho -  \Big( x_{0} + \dfrac{1}{2} \Big) \rho_{q} \bigg] 
+ \dfrac{1}{4} \Big( t_{1} + t_{2}+\dfrac{t_{1}x_{1}+t_{2}x_{2}}{2} \Big) \tau \nonumber \\ 
&&+ \dfrac{1}{8} \Big( t_{2} - 3t_{1}-\dfrac{3t_{1}x_{1}}{2} +\dfrac{t_{2}x_{2}}{2} \Big) \nabla^{2} \rho 
+ \dfrac{1}{16} (t_{2} + 3t_{1} + 6t_{1}x_{1}+2t_{2}x_{2}) \nabla^{2} \rho_{q} \nonumber \\
&&+ \dfrac{1}{8} (t_{2} - t_{1}+t_{2}x_{2}-t_{1}x_{1}) \tau_{q} - \dfrac{W_{0}}{2}  \nabla \cdot(\pmb{J} + \pmb{J}_{q} ) \nonumber \\ 
&&+ \dfrac{t_{3}}{12} \rho^{\alpha -1 } \Big[ (\alpha +2) \Big(1+\dfrac{x_{3}}{2} \Big) \rho^{\alpha +1} - \Big(x_{3}+\dfrac{1}{2}\Big)(\alpha \rho^{\alpha-1} \rho_{q}^{2} + 2 \rho^{\alpha} \rho_{q}) - x_{3} \alpha \rho^{\alpha -1} \pmb{s}^{2} \Big] \\
\label{eq:spino}\pmb{B}_{q} \ &=& \ \dfrac{W_{0}}{2} \pmb{\nabla} (\rho + \rho_{q})\\
\label{eq:gamma} \gamma_{q, \mu \nu} \ &=& \ - \dfrac{1}{4} (t_{2} - t_{1}) J_{q, \mu \nu} - \dfrac{t_{1}x_{1}+t_{2}x_{2}}{4}J_{\mu \nu} + \dfrac{1}{2} \Bigg[ (t_{e} +t_{o}) J_{ \mu \nu} - (t_{e} -t_{o})J_{q, \mu \nu} \Bigg] \nonumber \\ 
&&- \dfrac{3}{4} \Bigg[ (t_{e} + t_{o} ) J_{ \nu \mu} - (t_{e} - t_{o}) J_{q, \nu \mu} \Bigg]
 - \dfrac{3}{4} \Bigg[ (t_{e}+ t_{o}) J_{ \mu \nu} \delta_{\mu \nu}  - (t_{e} - t_{o}) J_{q, \mu \nu} \delta_{\mu \nu} \Bigg]\\
\pmb{I}_{q} \ &=& \ - \dfrac{1}{2} \Big( t_{1} + t_{2} + \dfrac{t_{1}x_{1}+t_{2}x_{2}}{2} \Big) \pmb{j} - \dfrac{1}{4} (t_{2} - t_{1}+2t_{2}x_{2}-2t_{1}x_{1}) \pmb{j}_{q} - \dfrac{W_{0}}{2} \big( \nabla \times (\pmb{s} + \pmb{s}_{q}) \big)\\
\pmb{C}_{q} \ &=& \ \dfrac{1}{8} (t_{2} - t_{1}) \pmb{s}_{q} +\dfrac{t_{1}x_{1}+t_{2}x_{2}}{8} \pmb{s} - \dfrac{1}{4} \Bigg[ (t_{e} + t_{o}) \pmb{s}(r) - (t_{e} - t_{o}) \pmb{s}_{q} (r) \Bigg]\\
\pmb{\Sigma}_{q} \ &=& \ \dfrac{t_{0}}{2} (x_{0} \pmb{s} - \pmb{s}_{q}) + \dfrac{1}{16} (3t_{1} + t_{2} ) \nabla^{2} \pmb{s}_{q} + (3t_{2}x_{2}-3t_{1}x_{1})\nabla^{2} \pmb{s} \nonumber \\
&&+ \dfrac{t_{1}x_{1}+t_{2}x_{2}}{8} \pmb{T} + \dfrac{1}{8} (t_{2} - t_{1} ) \pmb{T}_{q} +\dfrac{t_{3}}{6} (x_{3}\rho^{\alpha}\pmb{s} -\rho^{\alpha}\pmb{s}_{q}) - \dfrac{W_{0}}{2} \Big( \nabla \times (\pmb{j} + \pmb{j}_{q} )\Big) \nonumber \\ 
&&- \dfrac{3}{8} \Bigg[ (3t_{e} - t_{o}) \nabla \bigg( \nabla \cdot \pmb{s} (r) \bigg) - (3t_{e} + t_{o})  \nabla \bigg( \nabla \cdot \pmb{s}_{q} \bigg) \Bigg] - \dfrac{1}{4} \Bigg[ (t_{e} + t_{o}) \pmb{T} (r) - (t_{e} - t_{o}) \pmb{T}_{q} (r) \Bigg] \nonumber  \\
&&+ \dfrac{3}{4} \Bigg[ (t_{e} + t_{o}) \pmb{F} (r) - (t_{e} - t_{o}) \pmb{F}_{q} (r) \Bigg] + \dfrac{1}{8} \Bigg[ (3t_{e} - t_{o}) \nabla^{2} \pmb{s} (r) - (3t_{e} + t_{o}) \nabla^{2} \pmb{s}_{q} (r)  \Bigg]\\
\label{eq:dterm} D_{q, \mu} \ &=& \ \dfrac{3}{4} \Bigg[ (t_{e} + t_{o}) s_{\mu} (r) - (t_{e} - t_{o}) s_{q,\mu} (r) \Bigg]. 
\eea

\bibliographystyle{apsrev4-1}
\biboptions{sort&compress}

\section*{Bibliography}
\bibliography{pds-bib}
\end{document}